\documentclass[preprint,amsmath,amssymb,prb,aps,showpacs]{revtex4}
\usepackage{graphicx}
\usepackage{textcomp}
\usepackage{amsmath,amssymb}
\usepackage{mathrsfs}
\usepackage{color}
\usepackage{afterpage}
\usepackage{bm}
\usepackage[version=3]{mhchem}
\usepackage[normal]{subfigure}
\usepackage{enumitem}
\usepackage{hhline}
\usepackage{listings}
\usepackage[normalem]{ulem}

\newcommand{\R}{\mathbb{R}}

\newcommand\T{\rule{0pt}{2.6ex}}       

\begin{document}
\title{Learning physical descriptors for materials science\\ by compressed 
sensing}

\author{Luca M. Ghiringhelli$^1$, Jan Vybiral$^2$, Emre Ahmetcik$^1$, Runhai 
Ouyang$^1$, Sergey V. Levchenko$^1$, Claudia Draxl$^{3,1}$, and Matthias 
Scheffler$^{1,4}$}
\affiliation{$\phantom{ }^1$ Fritz-Haber-Institut der Max-Planck-Gesellschaft, 
Berlin-Dahlem, Germany \\
$\phantom{ }^2$ Charles University, Department of Mathematical Analysis, Prague, 
Czech Republic \\
$\phantom{ }^3$ Humboldt-Universit\"{a}t zu Berlin, Institut f\"{u}r Physik and 
IRIS Adlershof, Berlin, Germany\\
$\phantom{ }^4$ Department of Chemistry and Biochemistry and Materials 
Department, University of California--Santa Barbara, Santa Barbara, CA 
93106-5050, USA
}

\date{\today}
\begin{abstract}
The availability of big data in materials science offers new routes for 
analyzing materials properties and functions and achieving scientific 
understanding. Finding structure in these data that is not directly visible by 
standard tools and exploitation of the scientific information
requires new and dedicated methodology based on approaches from statistical 
learning, compressed sensing, and other recent methods from applied mathematics, 
computer science, statistics, signal processing, and information science. 
In this paper, we explain and demonstrate a compressed-sensing based methodology 
for feature selection, specifically for discovering physical descriptors, i.e., 
physical parameters that describe the material and its properties of interest, 
and associated equations that explicitly and quantitatively describe those 
relevant properties. As showcase application and proof of concept, we describe 
how to build a physical model for the quantitative prediction of the crystal 
structure of binary compound semiconductors.
\end{abstract}
\maketitle

\section{Introduction}

Big-data-driven research offers new routes towards scientific insight.
This big-data challenge is not only about storing and processing huge amounts of
data, but also, and in particular, it is a chance for new methodology for 
modeling, describing, and understanding.
Let us explain this by a simple, but most influential, historic example. About 
150 years ago, D. Mendeleev and others organized the 56 atoms that were known at 
their time in terms of a table, according to their weight and chemical
properties. Many atoms were not yet identified, but from the table, it was clear 
that they should
exist, and even their chemical properties were anticipated. The scientific 
reason behind the table,
i.e., the shell structure of the electrons, was unclear and was understood only 
50 years later
when quantum mechanics had been developed. Finding structure in the huge space 
of materials,
e.g., one table (or a map) for each property, or function of interest, is one of 
the great dreams of
(big-)data-driven materials science.

Obviously, the challenge to sort all materials is much bigger than that of the 
mentioned example of the Periodic Table of Elements (PTE). 
Up to date, the PTE contains 118 elements that have been observed.
About $200\, 000$ materials are ``known'' to exist \cite{LB}, but only for very 
few of these ``known'' materials, 
the basic properties (elastic constants, plasticity, piezoelectric 
tensors, thermal and electrical conductivity, 
etc.) have been determined. 
When considering surfaces, interfaces, nanostructures, organic materials, and 
hybrids of these mentioned systems, the amount of possible materials is 
practically infinite. 
It is, therefore, highly likely that new materials with superior and up to now 
simply unknown characterstics exist that could
help solving fundamental issues in the fields of energy, mobility, safety, 
information, and health.

For materials science it is already clear that big data are
structured, and in terms of materials properties and functions, the space of all 
possible materials is sparsely populated. Finding this structure in the big data, e.g., asking for 
efficient catalysts for CO$_2$ activation or oxidative coupling of
methane, good thermal barrier coatings, shape memory alloys as artery stents, or
thermoelectric materials for energy harvesting from temperature gradients, just 
to name a few examples, may be possible, even if the actuating physical mechanisms of these 
properties and functions are not yet understood in detail. Novel big-data analytics tools, 
e.g., based on machine learning or compressed sensing, promise to do so.

Machine learning of big data has been used extensively in a variety of fields 
ranging from biophysics and drug design to social media and text mining. It is typically considered a 
universal approach for ‘learning’ (fitting) a complex relationship $y=f(x)$. Some, though few, 
machine-learning based works have been done in materials science, e.g., Refs. 
\onlinecite{LorenzNN04, Ceder10, BehelrRev11, Ramprasad13, Csanyi2014, 
Mueller2014, Gross2014, Tanaka2015, VonLil2015, DeVita2015, Curtarolo15, 
Pilania2016-1, Pilania2016-2, COMBO2016, VonLil2016, Hammerschmidt16}. Most of 
them use kernel ridge regression or Gaussian processes, where in both cases the 
fitted/learned property is expressed as a weighted sum over all or selected data 
points. A clear breakthrough, demonstrating a `game change', is missing, so far, 
and a key problem, emphasized in Ref. \onlinecite{Ghiringhelli2015}, is that a 
good feature selection for descriptor identification is central to achieving a 
good-quality, predictive statistically learned equation. 

One of the purposes of discovering structure in materials-science data, is to 
predict a property of interest for a given complete class of materials. In order 
to do this, a model needs to be built, that maps some accessible, descriptive 
input quantities, identifying a material, into the property of interest. In 
statistical learning, this set of input parameters is called descriptor 
\cite{Ghiringhelli2015}. 
For our materials science study, this descriptor (elsewhere also termed ``fingerprint'' 
\cite{Ramprasad13,Curtarolo15}) is a set of physical parameters that uniquely describe the material and its function of interest.
It is important that materials that are very 
different (similar) with respect to the property or properties of interest, 
should be characterized by very different (similar) values of the descriptors. 
Moreover, the determination of the descriptor must not involve 
calculations as intensive as those needed for the evaluation of the property to 
be predicted.

In most papers published so far, the descriptor has been introduced \emph{ad 
hoc}, i.e., without performing an extensive, systematic analysis and without 
demonstrating that it is the best (in some sense) within a certain broad class.
The risk of such \emph{ad hoc} approach is that the learning step is inefficient 
(i.e., very many data points are needed), the governing physical mechanisms 
remain hidden, and the reliability of predictions is unclear. Indeed, machine 
learning is understood as an interpolation and not extrapolation approach, and 
it will do so well, if enough data are available. However, quantum mechanics and 
materials science are so multifaceted and intricate that we are convinced that 
we hardly ever will have enough data. Thus, it is crucial to develop these tools 
as domain-specific methods and to introduce some scientific understanding, 
keeping the bias as low as possible.

In this paper, we present a methodology for discovering physically meaningful 
descriptors and for predicting physically (materials-science) relevant 
properties. The paper is organized as follows: In section II, we introduce 
compressed-sensing\cite{Donoho06, Candes08, Kutyniok12, Vybiral15} based methods 
for feature selection; in section III, we introduce efficient algorithms for the construction and 
screening of feature spaces in order to single out ``the best'' descriptor. ``Feature selection'' 
is a widespread set of techniques that are used in statistical analysis in 
different fields \cite{FitSel}. In the following sections, we analyze the 
significance of the descriptors found by the statistical-learning methods, by 
discussing the interpolation ability of the model based on the found descriptors 
(section IV), the stability of the models in terms of sensitivity analysis 
(section V), and the extrapolation capability, i.e., the possibility of 
predicting new materials (section VI).

As showcase application and proof of concept, we use the quantitative prediction 
of the crystal structure of binary compound semiconductors, which are known to 
crystallize in zincblende (ZB), wurtzite (WZ), or rocksalt (RS) structures. The 
structures and energies of ZB and WZ are very close and for the sake of clarity 
we will not distinguish them here. 
For many materials, the energy difference between ZB and RS is larger, though 
still very small, namely just 0.001{\%} or less of the total energy of a single 
atom. Thus, high accuracy is required to predict this difference. The property 
$P$ that we aim to predict is the difference in the energies between RS and ZB 
for the given AB-compound material, $\Delta E_\textrm{AB}$. The energies are 
calculated within the Kohn-Sham formulation\cite{KohnSham65} of 
density-functional theory \cite{DFT64}, with the local-density approximation 
(LDA) exchange-correlation functional \cite{CA80,PW92}. Since in this paper we 
are concerned  with the development of a data-analytics approach, it is not 
relevant that also approximations better than LDA exist, only the internal consistency and the realistic complexity of the data are important. 
The below described methodology applies to these better approximations as well.

Clearly, the sign of $\Delta E_\textrm{AB}$ gives the classification (negative for RS and positive for ZB), but the 
quantitative prediction of $\Delta E_\textrm{AB}$ values gives a better 
understanding. The separation of RS and ZB structures into distinct classes, on 
the basis of their chemical formula only, is difficult and has existed as a key 
problem in materials science for over 50 years 
\cite{vanVechten69, Phillips70, Bloch72, Bloch74, Zunger80, Chelikowsky82, 
Pettifor84, Villars85, Tosi85, Galli87, Chelikowski12}. 
In the rest of the paper, we will always write the symbol $\Delta E$, without the subscript AB, to signify 
the energy of the RS compound minus the energy of the ZB compound.

\section{Compressed-sensing based methods for feature selection}

Let us assume that we are given data points $\{\bm{d}_1, P_1\},\dots,\{\bm{d}_N, 
P_N\}$, with $\bm{d}_1,\dots,\bm{d}_N\in\R^M$ and $P_1,\dots,P_N\in\R.$ We would 
like to analyze and understand the dependence of the output $P$ (for ``Property'') 
on the input $\bm{d}$ (for ``descriptor''), that reflects the 
specific measurement or calculation. Specifically, we are seeking for an {\em 
interpretable} equation $P=f(\bm{d})$, e.g., an explicit analytical 
formula, where $P$ depends on a small number of input parameters, i.e., the 
dimension of the input vector $\bm{d}$ is small. 
Below, we will impose constraints in order to arrive at a method for 
finding a low-dimensional linear solution for $P=f(\bm{d})$. Later, 
we introduce nonlinear transformations of the vector $\bm{d}$; dealing with such 
nonlinearities is the main focus of this paper.

When looking for a linear dependence $P=f(\bm{d})=\bm{d}\cdot\bm{c}$, the most
simple approach to the problem is the method of \emph{least squares}. This is 
the solution of
\begin{equation}\label{eq:ls}
\mathop{\rm arg min}\limits_{\bm{c} \in \R^M} \sum_{j=1}^N\Bigl(P_j-\sum_{k=1}^M 
d_{j,k} c_k\Bigr)^2= \mathop{\rm arg min}\limits_{\bm{c} \in \R^M} 
\|\bm{P}-\bm{D}\bm{c}\|_2^2,
\end{equation}
where $\bm{P}=(P_1,\dots,P_N)^T$ is the column vector of the outputs and 
$\bm{D}=(d_{j,k})$ is the $(N\times M)$-dimensional matrix of the inputs, and 
the column vector $\bm{c}$ is to be determined. The equality defines the 
$\ell_2$ or Euclidean norm ($\|\cdot\|_2$) of the vector $\bm{P}-\bm{D}\bm{c}$. 
The resulting function would then be just the linear relationship 
$f(\bm{d})=\langle \bm{d}, \bm{c} \rangle = \sum_{k=1}^M d_k c_k$. 
It is the function that minimizes the error $ 
\|\bm{P}-\bm{D}\bm{c}\|_2^2$ among all linear functions of $\bm{d}$. Let us 
point out a few properties of Eq. \eqref{eq:ls}.
First, the solution of Eq. \eqref{eq:ls} is given by an explicit formula 
$\bm{c}=(\bm{D}^T\bm{D})^{-1}\bm{D}^T\bm{P}$ if $\bm{D}^T\bm{D}$ is a 
non-singular matrix. Second, it is a convex problem, therefore the solution 
could also be found efficiently even in large dimensions (i.e. if $M$ and $N$ 
are large) and there are many solvers available \cite{nonlinpro}.
Finally, it does not put any more constraints on $\bm{c}$ (or $f$) than that it 
is linear and minimizes Eq. \eqref{eq:ls}.

A general linear function $f$ of $\bm{d}$ usually involves also an absolute 
term, i.e. it has a form $f(\bm{d})=c_0+\sum_{k=1}^M d_k c_k$.
It is easy to add the absolute term (also called \emph{bias}) to Eq. 
\eqref{eq:ls} in the form
\begin{equation}\label{eq:ls_bias}
\mathop{\rm arg min}\limits_{c_0\in\R,\bm{c}\in \R^{M}} \sum_{j=1}^N\Bigl(P_j- 
c_0-\sum_{k=1}^M d_{j,k} c_k\Bigr)^2.
\end{equation}
This actually coincides with Eq. \eqref{eq:ls} if we first enlarge the matrix 
$\bm{D}$ by adding a column full of ones. We will tacitly assume in the 
following that this modification was included in the implementation and shall 
not repeat that any more.

If pre-knowledge is available, it can be (under some conditions) incorporated 
into Eq. \eqref{eq:ls}. For example, we might want to prefer
those vectors $\bm{c}$, which are small (in some sense). This gives rise to 
\emph{regularized} problems. In particular, the \emph{ridge regression}
\begin{equation}\label{eq:rr}
\mathop{\rm arg min}\limits_{\bm{c}\in \R^M} \left( 
\|\bm{P}-\bm{D}\bm{c}\|_2^2+\lambda\|\bm{c}\|^2_2 \right)
  \end{equation}
with a penalty parameter $\lambda>0$ weights the magnitude of the vector 
$\bm{c}$ and of the error of the fit against each other. The larger $\lambda$,
the smaller is the minimizing vector $\bm{c}$. The smaller $\lambda$, the better 
is the least square fit (the first addend in Eq. \eqref{eq:rr}). More 
specifically, if $\lambda\to 0$, the solutions of Eq. \eqref{eq:rr} converge to 
the solution of Eq. \eqref{eq:ls}. And if $\lambda\to \infty$, the solutions of 
Eq. \eqref{eq:rr} tend to zero.

\subsection{Sparsity, NP-hard problems, and convexity of the minimization 
problem}

The kind of pre-knowledge, which we will discuss next, is that we would like 
$f(\bm{d})$ depend only on a small number of components of $\bm{d}$.
In the notation of Eq. \eqref{eq:ls}, we would like to achieve that most of the 
components of the solution $\bm{c}$ are zero.
Therefore, we denote the number of non-zero coordinates of a vector $\bm{c}$ by
\begin{equation}\label{eq:l0}
\|\bm{c}\|_0=\#\{j:c_j\not=0\}.
\end{equation}
Here, $\{j:c_j\not=0\}$ stands for the set of all the coordinates $j$, such that 
$c_j$ is non-zero, and $\#\{\ldots\}$ denotes
the number of elements of the set $\{\ldots\}$. Thus, $\|\bm{c}\|_0$ is the 
number of non-zero elements of the vector $\bm{c}$, and it is often called the 
$\ell_0$ norm of $\bm{c}$. We say that $\bm{c}\in\R^M$ is $k$-sparse, if 
$\|\bm{c}\|_0\le k$, i.e. if at most $k$ of its coordinates are not equal to 
zero.

When trying to regularize Eq. \eqref{eq:ls} in such a way that it prefers sparse 
solutions $\bm{c}$, one has to minimize: 
\begin{equation}\label{eq:rr0}
\mathop{\rm arg min}\limits_{\bm{c}\in \R^{M}} \left( 
\|\bm{P}-\bm{D}\bm{c}\|_2^2+\lambda\|\bm{c}\|_0 \right).
\end{equation}
This is a rigorous formulation of the problem that we want to solve, but it has a significant drawback: It 
is computationally infeasible when $M$ is large. This can be easily understood 
as follows. The naive way of solving Eq. \eqref{eq:rr0} is to look first on all 
index sets $T\subset\{1,\dots,M\}$ with one element and try to minimize over 
vectors supported in such sets. Then one looks for all subsets with two, three, 
and more elements. Unfortunately. their number grows quickly with $M$ and the 
level of sparsity $k$. It turns out, that this naive way can not be improved 
much. The problem is called ``NP-hard''\cite{Arora09}. In a ``Non-deterministic 
Polynomial-time hard'' problem, a good candidate for the solution can be checked 
in a polynomial time, but the solution itself can not be found in polynomial 
time. The basic reason is that Eq. \eqref{eq:rr0} is not convex.

\subsection{Methods based on the $\ell_1$ norm}

Reaching a compromise between convexity and promoting sparsity is made possible 
by the LASSO (Least Absolute Shrinkage and Selection Operator)\cite{LASSO} 
approach, in which the $\ell_0$ regularization of Eq. \eqref{eq:rr0} is replaced 
by the $\ell_1$ norm:
\begin{equation}\label{eq:rr1}
\mathop{\rm arg min}\limits_{\bm{c}\in \R^{M}} 
\|\bm{P}-\bm{D}\bm{c}\|_2^2+\lambda\|\bm{c}\|_1.
\end{equation}
The use of the $\ell_1$ norm ($\|\bm{c}\|_1=\sum_k|c_k|$), also known as the 
``Manhattan'' or ``Taxicab'' norm, is crucial here. On the one hand, the 
optimization problem is convex \cite{Tibshirani09}. On the other hand, the 
geometry of the $\ell_1$-unit ball
$\{\bm{x}\in\R^M:\|\bm{x}\|_1\le 1\}$ shows, that it promotes sparsity 
\cite{Kutyniok12,Vybiral15}.

Similarly to Eq. \eqref{eq:rr}, the larger the $\lambda>0$, the smaller the 
$\ell_1$-norm of the solution of Eq. \eqref{eq:rr1} would be, and vice versa.
Actually, there exists a smallest $\lambda_0 > 0$, such that the solution of Eq. 
\eqref{eq:rr1} is zero. If $\lambda$ then falls below this threshold, one or 
more components of $\bm{c}$ become non-zero. 

\subsection{Compressed sensing}

The minimization problem in Eq. \eqref{eq:rr1} is an approximation of the 
problem in Eq. \eqref{eq:rr0}, and their respective results do not necessarily 
coincide. The relation between these two minimization problems was studied 
intensively \cite{Greenshtein06,deGeer08,deGeer11} and we summarize the results, 
developed in the context of a recent theory of compressed sensing (CS) 
\cite{Donoho06,Kutyniok12,Vybiral15}. These will be useful for justifying the 
implementation of our method (see below).
The literature on CS is extensive and we believe that the following notes are a 
useful compendium.

We are especially interested in conditions on $\bm{P}$ and $\bm{D}$ which 
guarantee that the solutions
of Eqs. \eqref{eq:rr0} and \eqref{eq:rr1} coincide, or at least do not differ 
much.
We concentrate on a simplified setting, namely attempting to find 
$\bm{c}\in\R^M$ fulfilling $\bm{Dc}=\bm{P}$ with fewest non-zero entries.
This is written as 
\begin{equation}\label{eq:CS1}
\mathop{\rm arg min}\limits_{\bm{c}:\bm{D}\bm{c}=\bm{P}}\|\bm{c}\|_0 .
\end{equation}
As mentioned above, this minimization is computationally infeasible, and 
we are interested in finding the same solution by its convex 
reformulation:
\begin{equation}\label{eq:CS2}
\mathop{\rm arg min}\limits_{\bm{c}:\bm{D}\bm{c}=\bm{P}}\|\bm{c}\|_1.
\end{equation}
The answer can be characterized by the notion of the Null Space Property (NSP) 
\cite{NSP09}.
Let $\bm{D}\in \R^{M\times N}$ and let $\Omega\in\{1,\dots,N\}$. Then $\bm{D}$ 
is said to have the NSP of order $\Omega$ if
\begin{equation}\label{eq:def:NSP}
\sum_{j\in T}|v_j|<\sum_{j\not\in T}|v_{j}|\quad \forall \bm{v}\not=0 \ 
\text{with} \ \bm{D}\bm{v}=0\ \forall T\subset\{1,\dots,N\}\ 
\text{with}\ \#T\le \Omega.
\end{equation}
It is shown \cite{NSP09} that every $\Omega$-sparse vector $\bm{x}$ is the 
unique solution of Eq. \eqref{eq:CS2} with $\bm{P}=\bm{D}\bm{x}$ if, and only 
if, $\bm{D}$ has the NSP of order $\Omega$.
As a consequence, the minimization of the $\ell_1$ norm as given by Eq. 
\eqref{eq:CS2} recovers the unknown $\Omega$-sparse vectors ${\bm c}\in\R^N$
from $\bm{D}$ and $\bm{P}=\bm{D}\bm{c}$ only if $\bm{D}$ satisfies the NSP of 
order $\Omega$.
Unfortunately, when given a specific matrix $\bm{D}$, it is not easy to check if 
it indeed has the NSP.

However, the CS analysis gives us some guidance for the admissible dimension $M$ 
of $\bm{D}.$
It is known \cite{NSP09,Vybiral15} that there exists a constant $C>0$ such that 
whenever 
\begin{equation}\label{eq:CS3}
M\ge C \ \! \Omega \ \! \ln(N)
\end{equation}
then there exist matrices $\bm{D}\in \R^{M\times N}$
with NSP of order $\Omega$. Actually, a random matrix with these dimensions 
satisfies the NSP of order $\Omega$ with high probability.
On the other hand, this bound is tight, i.e., if $M$ falls below this 
bound no stable and robust recovery of $\bm{c}$ from $\bm{D}$ and $\bm{P}$ is possible, and this is true 
for every matrix $\bm{D}\in\R^{M\times N}.$
Later on, $M$ will correspond to the number of AB compounds considered and will 
be fixed to $M=82$. Furthermore,
$N$ will be the number of input features, and Eq. \eqref{eq:CS3} tells us that 
we can expect reasonable performance of
$\ell_1$-minimization only if their number is at most of the order 
$e^{\frac{M}{C\Omega}}.$

Furthermore, the CS analysis sets forth that the NSP is surely violated already for 
$\Omega=2$ if any two columns of $\bm{D}$ are a multiple of each other.
In general, one can expect the solution of Eq. \eqref{eq:CS2} to differ 
significantly from the solution of Eq. \eqref{eq:CS1}
if any two columns of $\bm{D}$ are highly correlated or, in general, if any of 
the columns of $\bm{D}$ lies close to the span
of a small number of any other its columns.

Historically, LASSO appeared first in 1996 as a machine-learning (ML) technique, 
able to provide stable solutions to underdetermined linear systems, thanks to 
the $\ell_1$ regularization.
Ten years later, the CS theory appeared, sharing with LASSO the concept of 
$\ell_1$ regularization, but the emphasis is on the reconstruction of a possibly 
noisy signal $\bm{c}$ from a minimal number of observations.
CS can be considered the theory of LASSO, in the sense that it gives conditions 
(on the matrix $\bm{D}$) on when it is reasonable to expect that the $\ell_1$ 
and $\ell_0$ regularizations coincide.
If the data and a, typically overcomplete, basis set are known, then LASSO is 
applied as a ML technique, where the task is to identify the smallest number of 
basis vectors that yield a model of a given accuracy.
If the setting is such that a minimal set of measurements should be performed to 
reconstruct a signal, as in quantum-state tomography \cite{Eisert10}, then CS 
tells how the data should be acquired.
As it will become clear in the following, our situation is somewhat 
intermediate, in the sense that we have the data (here, the energy of the 82 
materials), but we have a certain freedom to select the basis set for the model. 
Therefore, we use CS to justify the construction of the matrix $\bm{D}$, and we 
adopt LASSO as a mean to find the low-dimensional basis set.

\subsection{A simple LASSO example: the energy differences of crystal structures}

In this and following sections, we walk through the application of LASSO to a 
specific materials-science problem, in order 
to introduce step by step a systematic approach for finding simple analytical 
formulas, that are built from simple input parameters (the {\em primary 
features}), for approximating physical properties of interest. 
We aim at predicting $\Delta E_\textrm{AB}$, the difference in DFT-LDA energy 
between ZB and RS structures in a set of 82 octet binary semiconductors. 
Calculations were performed using the all-electron full-potential code FHI-aims 
\cite{Blum09} with highly accurate basis sets, $\bf k$-meshes, integration 
grids, and scaled ZORA \cite{scaledZORA} for the relativistic treatment.

The order of the two atoms in the chemical formula AB is chosen such 
that element A is the cation, i.e., it has the smallest Mulliken 
electronegativity, $\textrm{EN}\! =\! -(\textrm{IP}+\textrm{EA})/2$. 
IP and EA are the ionization potential and electron affinity of the free, 
isolated, spinless, and spherical symmetric atom. 
As noted, the calculation of the descriptor must involve less intensive 
calculations than those needed for the evaluation of the property to be 
predicted. Therefore, we consider only properties of the free atoms A and B, 
that build the binary material, and properties of the gas-phase dimers. 
In practice, we identified the following {\em primary features}, 
exemplified for atom A: the ionization potential IP(A), the electron affinity 
EA(A) \cite{IPEA}, H(A) and L(A), i.e., the energies of the highest-occupied and 
lowest-unoccupied Kohn-Sham (KS) levels, and $r_s$(\textrm{A}), 
$r_p$(\textrm{A}), and $r_d$(\textrm{A}), i.e., the radii where the radial 
probability density of the valence $s$, $p$, and $d$ orbitals are maximal. The 
same features were used for atom B. 
Clearly, these primary features were chosen because it is well known since long 
that the relative ionization potentials, the atomic radii, and 
$sp$-hybridization govern the bonding in these materials. Consequently, some 
authors\cite{Bloch73, Phillips78, Phillips79} already recognized that certain 
combinations of $r_s$ and $r_p$ -- called $r_\sigma$ and $r_\pi$ (see 
below) -- may be crucial for constructing a descriptor that predicts the RS/ZB 
classification. Note that just the sign of $\Delta E$ was predicted, 
while the model we present here targets at a quantitative prediction of $\Delta 
E$. In contrast to previous work, we analyze how LASSO will perform in this task 
and how much better the corresponding description is. 
We should also note that the selected set of atomic primary features contains 
redundant physical information, namely H(A) and L(A) contain similar information 
as IP(A) and EA(A). In particular H(A) $-$ L(A) is correlated, on physical 
grounds, to IP(A) $-$ EA(A). However, since the values of these two differences 
are not the same \cite{IPEA}, all four features were included. As it will be 
shown below and in particular in Appendix \ref{A:A}, the two pairs of features are not 
interchangeable.

We start with a simplified example, in order to show in an easily reproducible 
way the performance of LASSO in solving our problem. To this extent, we 
describe each compound AB by the vector of the following six quantities:
\begin{equation}
\label{eq:featspace6}
 \bm{d}_{\textrm{AB}}=(r_s(\textrm{A}), r_p(\textrm{A}), r_d(\textrm{A}), 
r_s(\textrm{B}), r_p(\textrm{B}), r_d(\textrm{B})). 
\end{equation}
All this data collected gives a $82\times 6$ matrix 
$\bm{D}$ -- one row for each compound. We standardize $\bm{D}$ to have 
zero mean and variance 1, i.e. subtract from each column its mean and divide by 
its standard deviation. Standardization of the data is common 
practice\cite{Tibshirani09} and aimed at controlling the numerical stability of 
the solution. However, when non-linear combinations of the features and 
cross-validation is involved, the standardization has to be performed carefully (see below). 
The energy differences are stored in a vector $\bm{P}\in\R^{82}.$

We are aiming at two goals:
\begin{itemize}
\item On the one hand, we would like to minimize the \emph{mean-squared error} 
(MSE), a  measure of the quality of the approximation, given by
\begin{equation}
 \frac{1}{N}\sum_{j=1}^{N}\Bigl(P_j-\sum_{k=1}^M d_{j,k} c_k\Bigr)^2. 
\end{equation}
In this tutorial example, the size $N$ of the dataset is 82 and, with the above 
choice of Eq. \eqref{eq:featspace6}, $M=6$.
\item On the other hand, we prefer sparse solutions $\bm{c}^\dag$ 
with small $\Omega$, as we like 
to explain the dependence of the energy difference $\bm{P}$ based on a 
low-dimensional (small $\Omega$) {\em descriptor} $\bm{d}^\dag$. 
\end{itemize}
These two tasks are obviously closely connected, and go against each other. The 
larger the coefficient $\lambda$ in Eq. \eqref{eq:rr1}, the more sparse the 
solution $\bm{c}$ will be; the smaller $\lambda$, the smaller the 
mean-squared error will be. The choice of $\lambda$ is at our disposal and let 
us weight sparsity of $\bm{c}$ against the size of the MSE.

In the following, we will rather report the {\em root mean square error}(RMSE) 
as a quality measure of a model. The reason is that the RMSE has 
the same units as the predicted quantities, therefore easier to understand in 
absolute terms. Specifically, when a sparse solution $\bm{c}^\dag$ of Eq. 
\eqref{eq:rr1} is mentioned, with $\Omega$ non-zero components, we report the 
RMSE:
\[
\sqrt{ \frac{1}{82} \sum_{j=1}^{82} \Bigl( P_j - \sum_{k=1}^\Omega 
d_{j,k}^\dag c_k^* \Bigr)^2}, 
\]
where $\bm{D}^\dag = (d_{j,k}^\dag)$ contains the columns corresponding to the 
non-zero components of $\bm{c}^\dag$ and $\bm{c}^* = (c_k^*)$ is the solution of 
the least square fit: 
\[
 \mathop{\rm arg min}\limits_{\bm{c}^*\in \R^{\Omega}} 
\|\bm{P}-\bm{D}^\dag\bm{c}^*\|_2^2. 
\]

Let us first state that the MSE obtained when setting $\bm{c}=0$ is 0.1979 
eV$^2$, which corresponds to a RMSE of 0.4449 eV. This means that if one 
``predicts'' $\Delta E=0$ for all the materials, the RMSE of such ``model'' is 
0.4449 eV. This number acts as a baseline for judging the quality of new models.

In order to run this and all the numerical tests in this paper, we have written 
{\sc Python} scripts and used the linear-algebra solvers, including the LASSO 
solver, from the scikit\_learn ({\tt sklearn}) library (see Appendix \ref{A:D} for the 
actual script used for this section). We like to calculate 
the coefficients for a decreasing sequence of a hundred 
$\lambda$ values, starting from the smallest $\lambda = \lambda_1$, such that 
the solution is $\bm{c}=0$. $\lambda_1$ is determined by $N \lambda_1 = 
\mathop{\rm max}\limits_{i} \langle \bm{d}_i \,, 
\bm{c}\rangle$~\cite{Tibshirani09}. The sequence of $\lambda$ values is 
constructed on a log scale, and the smallest $\lambda$ value is set to 
$\lambda_{100} = 0.001 \lambda_{1}$. In our case, $\lambda_{1}=0.3169$ and the 
first non-zero component of $\bm{c}$ is the second one, corresponding to 
$r_p(\textrm{A})$. 
At $\lambda_{19} = 0.0903$, the second non-zero 
component of $\bm{c}$ appears, corresponding to $r_d(\textrm{B})$ (see Eq. 
\eqref{eq:featspace6}). For decreasing $\lambda$, more and more components of 
$\bm{c}$ are different from zero until, from the column corresponding to 
$\lambda_{67}=0.0032$ down to $\lambda_{100}=0.0003$, all entries are 
occupied,  i.e. no sparsity is left. The method clearly suggests 
$r_p(\textrm{A})$ to be the most useful feature for predicting the energy 
difference. Indeed, by enumerating all six linear (least-square) models 
constructed with only one of the components of vector $\bm{d}$ at a time, the 
smallest RMSE is given by the model based on $r_p(\textrm{A})$.
We conclude that LASSO really finds the coordinate best describing (in a 
linear way) the dependence of $P$ on $\bm{d}$.

Let us now proceed to the best pair. The second appearing non-zero component is 
$r_d(B)$. The RMSE for the pair $(r_p(\textrm{A}), r_d(\textrm{B}))$ is 0.2927 
eV. An exhaustive search over all the fifteen pairs of elements of 
$\bm{d}_{\textrm{AB}}$ reveals that the pair $(r_p(\textrm{A}), 
r_d(\textrm{B}))$, indeed, yields the lowest (least-square) RMSE among 
all possible pairs in $\bm{d}_{\textrm{AB}}$.

The outcome for the best triplet reveals the weak point of the 
method. The third non-zero components of $\bm{d}$ is $r_p(\textrm{B})$, and the 
RMSE of the triplet $r_p(\textrm{A}),r_p(\textrm{B}),r_d(\textrm{B})$ is 0.2897 
eV, while an exhaustive search over all the triplets found, suggests 
($r_s(\textrm{B}), r_p(\textrm{A}), r_p(\textrm{B})$) to be optimal with a RMSE 
error of 0.2834, i.e., some 2\% better. The reason is that Eq. \eqref{eq:rr1} is 
only a convex proxy of the actual problem in Eq. \eqref{eq:rr0}. Their solutions 
have similar performance in terms of RMSE, but they do not have to coincide. Now 
let us compare the norm $\|\bm{c}\|_1$ of both triplets, for $\bm{c}^*$ obtained 
by the least square regression with standardized columns. For the optimal 
triplet, $(r_s(\textrm{B}), r_p(\textrm{A}), r_p(\textrm{B}))$, $\|\bm{c}\|_1 = 
3.006$  and for $(r_p(\textrm{A}),r_p(\textrm{B}),r_d(\textrm{B}))$, 
$\|\bm{c}\|_1 = 0.5843$. Here we observe, that the first one needs 
higher coefficients for a small $\|\bm{P}-\bm{D}\bm{c}\|_2^2$, while the 
second one provides a better compromise between a small 
$\|\bm{P}-\bm{D}\bm{c}\|_2^2$ and a small $\lambda\|\bm{c}\|_1$ in Eq. 
\eqref{eq:rr1}. The reason for the large coefficients for the former is 
that considering the 82-dimensional vectors whose components are the values of 
$r_s(\textrm{B})$ and $r_p(\textrm{B})$ for all the materials (in the same order), these vectors 
are almost parallel (their Pearson correlation coefficient is 0.996)\cite{CorrBerkson}. 
In order to understand this, let us have a look at both least square models:
\begin{align}
 \Delta E_{\ell_0} = + 1.272 r_s(\textrm{B}) - 0.296 r_p(\textrm{A}) - 1.333 
r_p(\textrm{B}) + 0.106\\
 \Delta E_{\ell_1} = - 0.337 r_p(\textrm{A}) - 0.044 r_p(\textrm{B}) - 0.097 
r_d(\textrm{B}) + 0.106. 
\end{align}
In the optimal $\ell_0$ model, the highly correlated vectors $r_s(\textrm{B})$ 
and $r_p(\textrm{B})$ appear approximatively as the difference $r_s(\textrm{B}) 
- r_p(\textrm{B})$ (their coefficients in the linear expansion have almost the 
same magnitude, $\sim 1.3$, and opposite sign). The difference of these two 
almost collinear vectors with the same magnitude (due to the applied 
standardization) is a vector much shorter than both $r_s(\textrm{B})$ and 
$r_p(\textrm{A})$, but also much shorter than $r_p(\textrm{B})$ and $P$, 
therefore $r_s(\textrm{B}) - r_p(\textrm{B})$ needs to be multiplied by a 
relatively large coefficient in order to be comparable with the other vectors. 
Indeed, while the coefficient of  $r_s(\textrm{B}) - r_p(\textrm{B})$ is about 
1.3, all the other coefficients, in particular in the $\ell_1$ solution, are 
much smaller.
Such large coefficient is penalized by the $\lambda\|\bm{c}\|_1$ term and 
therefore a sub-optimal (according to $\ell_0$-regularized minimization) triplet 
is selected. This examples shows how with highly correlated features, 
$\lambda\|\bm{c}\|_1$ may not be a good approximation for $\lambda\|\bm{c}\|_0$.

Repeating the LASSO procedure for the matrix $\bm{D}$ consisting once only of 
elements of the first triplet and once of the second triplet, shows that only 
for $\lambda \leq 0.00079$ the first triplet has a smaller LASSO-error, since 
then the contribution of the high coefficients to the LASSO-error are 
sufficiently small. When $\bm{D}$ contains all features, at this 
$\lambda$ already all coefficients are non-zero. 

For the sake of completeness, let us just mention that the RMSE for the pair of 
descriptors defined by John and Bloch \cite{Bloch74} as
\begin{eqnarray}
 r_\sigma &=& \big|\big(r_p(\textrm{A})+r_s(\textrm{A})\big)- 
\big(r_p(\textrm{B})+r_s(\textrm{B})\big) \big| \\
 r_\pi &=& 
\big|r_p(\textrm{A})-r_s(\textrm{A})\big|+\big|r_p(\textrm{A})-r_s(\textrm{A}
)\big|
\end{eqnarray}
is 0.3055 eV. For predicting the energy difference in a linear way, the 
pair of descriptors $(r_p(\textrm{A}),r_d(\textrm{B}))$ is already slightly 
better (by 4\%) than this.

\subsection{A more complex LASSO example: non-linear mapping}
Until this point, we have treated only linear models, i.e., where the function 
$P=f(\bm{d})$ is a linear combination of the components of the input vector 
$\bm{d}$. This is a clear limitation. 
In this section, we describe how one can easily introduce non-linearities, 
without loosing the simplicity of the linear-algebra solution. To the purpose, 
the vector $\bm{d}$ is mapped by a (in general non-linear) function 
$\Phi:\R^{M_1}\to \R^{M_2}$ into a higher-dimensional space, and only then the 
linear methods are applied in $\R^{M_2}$. This idea is well known in 
$\ell_2$-regularized {\em kernel} methods \cite{Hofmann08}, where the so-called 
kernel trick is exploited.
In our case, we aim at explicitly defining and evaluating a higher 
dimensional-mapping of the initial features, where each new dimension is a 
non-linear function of one or more initial features.

We stay with the case of 82 compounds. To keep the presentation simple, we leave 
out the inputs $r_d(\textrm{A})$ and $r_d(B)$.
Hence, every compound is first described by the following four {\em primary 
features},
\begin{equation}
 \bm{d}_{\textrm{AB}}=(r_s(\textrm{A}),r_p(\textrm{A}),r_s(B),r_p(B)). 
\end{equation}
We will construct a simple, but non-trivial non-linear mapping $\Phi:\R^4\to 
\R^{M^*}$ (with $M^*$ to be defined) and apply LASSO afterwards. The 
construction of $\Phi$ involves some of our pre-knowledge, for example, on 
dimensional grounds, we expect that expressions like 
$r_s(\textrm{A})^2-r_p(\textrm{A})$ have no physical meaning and should be 
avoided. Therefore, in practice, we allow for sums and differences only 
of quantities with the same units.
We hence consider the 46 features listed in Table 
\ref{T:featspace_small}: the 4 primary plus 42 {\em derived} features, building 
$\Phi(\bm{d}_{AB})$, represented by the matrix $\bm{D}$.

\begin{table}[h!]
\begin{tabular}{|c|l|c|}
\hline
Columns of $\bm{D}$ & Description & Typical formula\\
\hline
1-4 & Primary features & $r_s(\textrm{A}),r_p(\textrm{A}),r_s(B),r_p(B)$\\
5-16 & All ratios of all pairs of $r$'s & $r_s(\textrm{A})/r_p(\textrm{A})$\\
17-22 & Differences of pairs & $r_s(\textrm{A})-r_p(\textrm{A})$\\
23-34 & All differences divided by the remaining $r$'s & 
$(r_s(\textrm{A})-r_p(\textrm{A}))/r_s(B)$\\
35-40 & Absolute values of differences & $|r_s(\textrm{A})-r_p(\textrm{A})|$\\
41-43 & Sums of absolute values of differences &\\
& with no $r$ appearing twice & 
$|r_s(\textrm{A})-r_p(\textrm{A})|+|r_s(B)-r_p(B)|$\\
44-46 & Absolute values of sums of differences &\\
& with no $r$ appearing twice & 
$|r_s(\textrm{A})-r_p(\textrm{A})+r_s(B)-r_p(B)|$\\
\hline
\end{tabular}
\caption{Definition of the feature space for the tutorial example described in 
section II.E.
\label{T:featspace_small} }
\end{table}

The descriptors of John and Bloch \cite{Bloch74}, $r_\pi$ and $r_\sigma$, are 
both included in this set, with indexes 41 and 46, respectively.
The data are standardized, so that each of the 46 columns of $\bm{D}$ has mean 
zero and variance 1. Note that for columns 5-46, the standardization is applied 
only after the analytical function of the primary features is evaluated, i.e., 
for physical consistency, the primary features enter the formula unshifted and 
unscaled.

\begin{table}[h!]
\begin{tabular}{|c|c|c|c|}
\hline
$\lambda$ & \parbox[c]{1.1cm} {$\ell_0$} & Feature & Action \\ 
\hline
\parbox[c]{1.1cm} {0.317} & 1 & $r_p(\textrm{A})$ & + \\
0.257 & 2 & 
$|r_p(\textrm{A})-r_s(\textrm{A})|+|r_p(\textrm{B})-r_s(\textrm{B})|$ & + \\
0.240 & 3 & $r_s(\textrm{B})/r_p(\textrm{A})$ & + \\
0.147 & 4 & $r_s(\textrm{A})/r_p(\textrm{A})$ & + \\
0.119 & 3 & 
$|r_p(\textrm{A})-r_s(\textrm{A})|+|r_p(\textrm{B})-r_s(\textrm{B})|$ & $-$ \\
0.111 & 4 & $r_s(\textrm{A})$ & + \\
0.104 & 3 & $r_p(\textrm{A})$ & $-$ \\
0.084 & 4 & $|r_s(\textrm{B})-r_p(\textrm{B})|$ & + \\
0.045 & 5 & $(r_s(\textrm{B})-r_p(\textrm{B}))/r_p(\textrm{A})$ & + \\
0.034 & 6 & $|r_s(\textrm{A})-r_p(\textrm{B})|$ & + \\
0.028 & 5 & $r_s(\textrm{A})$ & $-$ \\
\hline
\end{tabular}
\caption{Bookkeeping of (decreasing) $\lambda$ values at which either a new 
feature gets non-zero coefficients (marked by a `+' in column Action) or a 
feature passes from non-zero to zero coefficient (marked by a `$-$'). The value 
of $\ell_0$ counts the non-zero coefficients at each reported $\lambda$ value. 
In a rather non-intuitive fashion, the number of non-zero coefficients 
fluctuates with decreasing $\lambda$, rather than monotonically increasing; this 
is an effect of linear correlations in the feature space.
\label{T:Bookkeeping}
}
\end{table}

Applying LASSO gives the result shown in Table \ref{T:Bookkeeping}, where we 
list the coordinates as they appear when $\lambda$ decreases. 
Where we have truncated the list, at $\lambda = 0.028$, the features 
with non-zero coefficient are:
\[
1: \frac{r_s(\textrm{A})}{r_p(\textrm{A})}, 2: 
\frac{r_s(\textrm{B})}{r_p(\textrm{A})}, 3: 
\frac{r_s(\textrm{B})-r_p(\textrm{B})}{r_p(\textrm{A})}, 4: 
|r_s(\textrm{A})-r_p(\textrm{B})|\ \text{and}\ 5: 
|r_s(\textrm{B})-r_p(\textrm{B})|.
\]
Let us remark, that the features 2 and 3 and the features 3 and 5 are strongly 
correlated, with covariance greater than 0.8 for both pairs \cite{transcorr}. 
This results in a difficulty for LASSO to identify the right features.

With these five descriptor candidates, we ran an exhaustive $\ell_0$ 
test over all the 5 $\cdot$ 4/2=10 pairs.
We discovered that the pair of the second and third selected features, i.e. of
\begin{equation}\label{eq:pair}
\frac{r_s(\textrm{B})}{r_p(\textrm{A})},\quad 
\frac{r_s(\textrm{B})-r_p(\textrm{B})}{r_p(\textrm{A})}
\end{equation}
achieves the smallest RMSE of $0.1520$ eV, 
improving on John and Bloch's descriptors \cite{Bloch74} by a factor 2. 
We also ran an exhaustive $\ell_0$ search for the optimal pair over all 
8 features that were singled out in the LASSO sweep, i.e., including also 
$r_p(\textrm{A})$, $r_\pi$, and $r_s(\textrm{A})$. The best pair was still the 
one in Eq. \eqref{eq:pair}.
We then also performed a full search over the 46~$\cdot$~45/2 = 1035 pairs, 
where the pair in Eq. \eqref{eq:pair} still turned out to be the one yielding 
the lowest RMSE.
We conclude that even though LASSO is not able to find directly the optimal 
$\Omega$-dimensional descriptor, it can efficiently be used for filtering the 
feature space and single out the ``most important'' features, whereas the 
optimal descriptor is then identified by enumeration over the subset identified 
by LASSO. 

The numerical test discussed in this section shows the following:
\begin{itemize}
 \item A promising strategy is to build an even larger feature space, by 
combining the {\em primary features} via analytic formulas. In the next section, 
we walk through this strategy for systematically constructing a large 
feature space.
 \item LASSO cannot always find the best $\Omega$-dimensional descriptor as the 
first $\Omega$ columns of $\bm{D}$ with non-zero coefficient by decreasing 
$\lambda$. This is understood as caused by features that are (nearly) linearly 
correlated. However, an efficient strategy emerges: First using LASSO for extracting a 
number $\Theta > \Omega $ of ``relevant'' components. These are the first $\Theta$ columns of $\bm{D}$ 
with non-zero coefficients found when decreasing $\lambda$. Then, performing an exhaustive search over all the 
$\Omega$-tuples that are subsets of the $\Theta$ extracted columns. The latter is in practice the problem formulated in Eqs. \eqref{eq:l0} and \eqref{eq:rr0}.
In the next section, we formalize this strategy, which we call henceforth 
LASSO+$\ell_0$, because it combines LASSO ($\ell_1$) and $\ell_0$ optimization.
\end{itemize}
 The feature-space construction and LASSO+$\ell_0$ strategies presented in the 
next section, are essentially those employed in our previous paper 
\cite{Ghiringhelli2015}. The purpose of this extended presentation is to 
describe a general approach for the solution of diverse problems, where the only 
requisite is that the set of basic ingredients (the {\em primary features}) is 
known. As a concluding remark of this section, we note that compressed sensing 
and LASSO were successfully demonstrated to help solving quantum-mechanics and 
materials-science problems in Refs. \onlinecite{Hart13,Ozolins13,OzolinsPNAS13,Ozolins14,Eisert14}. In all those 
papers, a $\ell_1$ based optimization was adopted to select from a well defined 
set of functions, in some sense the ``natural basis set'' for the specific 
problem, a minimal subset of ``modes'' that maximally contribute to the accurate 
approximation of the property under consideration. In our case, the application 
of the $\ell_1$ (and subsequent $\ell_0$) optimization must be preceded by the 
construction of a basis set, or feature space, for which a construction strategy 
is not at all {\em a priori} evident.\\
 
In the discussed numerical tests until this point, we have always looked for the 
low-dimensional model that minimizes the square error (the square of the 
$\ell_2$ norm of the fitting function). 
Another quantity of physical relevance that one may want to minimize is the 
maximum absolute error (maxAE) of the fit. This is called infinity norm and, for 
the vector $\bm{x}$, it is written as  $\|\bm{x}\|_\infty= \max |x_k|$. The 
minimization problem of Eq. \eqref{eq:rr1} then becomes
\begin{equation}\label{eq:rrinf}
\mathop{\rm arg min}\limits_{\bm{c}\in \R^{M}} 
\|\bm{P}-\bm{D}\bm{c}\|_\infty+\lambda\|\bm{c}\|_1 \ .
\end{equation}
This is still a convex problem. 
of Eq. \eqref{eq:rr1}.
We have looked for the model that gives the lowest MaxAE, starting from the 
feature space of size 46 defined in Table \ref{T:featspace_small}.

\begin{table}[h!]
\begin{tabular}{|c|c|c|c|c|}
\hline
\parbox[c]{1.1cm} {$\Omega$} & \parbox[c]{1.1cm} {Type} & \parbox[c]{1.4cm} 
{RMSE}& \parbox[c]{1.4cm} {MaxAE}  & Descriptor \\ 
\hline
1  & $\ell_2$ & 0.32 & 1.93 & $r_p(\textrm{A})$\\
1 & $\ell_\infty$ & 0.36 & 1.70 & $r_s(\textrm{A})/r_s(\textrm{B})$\\
\hline
2 & $\ell_2$ & 0.15 & 0.42 & $r_s(\textrm{B})/r_p(\textrm{A}), 
(r_s(\textrm{B})-r_p(\textrm{B}))/r_p(\textrm{A})$ \\
2 & $\ell_\infty$ & 0.15 & 0.42 & $r_s(\textrm{B})/r_p(\textrm{A}), 
(r_s(\textrm{B})-r_p(\textrm{B}))/r_p(\textrm{A})$\\
\hline
\end{tabular}
\caption{Comparison of descriptors selected by minimizing the $\ell_2$-norm of 
the fitting function (the usual LASSO problem) or the $\ell_\infty$ 
(maximum norm) over the feature space described in Table 
\ref{T:featspace_small}. Reported is also the performance of the models in terms 
of RMSE and MaxAE.
\label{T:l2lmax}}
\end{table}

For the specific example presented here, we find (see Table 
\ref{T:l2lmax}) that the 2D model is the same for both settings, i.e., the model 
that minimizes the RMSE also minimizes the MaxAE. This is, of course,not 
necessarily true in general. In fact, the 1D model that minimizes the RMSE 
differs from the 1D model that minimizes the MaxAE.

\section{Generation of a feature space}

For the systematic construction of the feature space, we first divide the {\em 
primary features} in groups, according to their physical meaning. In particular, 
necessary condition is that elements of each group are expressed with the same 
units.
We start from atomic features, see Table \ref{T:featspace0}. 

\begin{table}[h!]
\begin{tabular}{|c|l|l|c|}
\hline
\parbox[c]{1.cm} {ID} & Description & Symbols & \parbox[c]{1.cm} {\#} \T\\
\hline
$A1$ & Ionization Potential (IP) and Electron Affinity (EA) & IP(A), EA(A), 
IP(B), EA(B) & 4 \T\\
$A2$ & Highest occupied (H) and lowest unoccupied (L) & H(A), L(A), H(B), L(B) & 
4 \T\\
     & Kohn-Sham levels &  & \\
$A3$ & Radius at the max. value of $s$, $p$, and $d$ & $r_s(\textrm{A})$, 
$r_p(\textrm{A})$, $r_d(\textrm{A})$ & 6 \T\\
     & valence radial radial probability density & $r_s(\textrm{B})$, 
$r_p(\textrm{B})$, $r_d(\textrm{B})$ & \\
 \hline
\end{tabular}
\caption{Set of atomic primary features used for constructing the feature 
space, divided in groups. The `ID' labels the group and `\#' indicates the 
number of features in the group. 
\label{T:featspace0}}
\end{table}
Next, we define the combination rules. We note here that building algebraic 
function over a set of input variables (in our case, the {\em primary features}) 
by using a defined dictionary of algebraic (unary and binary) operators and 
finding the optimal function with respect to a given cost functional is the 
strategy of {\em symbolic regression} \cite{Koza98}. In this field of 
statistical learning, the optimal algebraic function is searched via an 
evolutionary algorithm, where the analytic expression is evolved by replacing 
parts of the test functions with more complex functions.
In other words, in symbolic regression, the evolutionary algorithm guides the 
construction of the algebraic functions of the primary features. In our case, we 
borrow from symbolic regression the idea of constructing functions by combining 
`building blocks' in more and more complex way, but the selection of the optimal 
function (in our language, the descriptor) is determined by the LASSO+$\ell_0$ 
algorithm over the whole set of generated functions, {\em after} the set of 
candidate functions is generated.

Our goal is to create ``grammatically correct'' 
combinations of the {\em primary features}. This means, besides applying 
the usual syntactic rule of algebra, we add a physically motivated constraint, 
i.e., we exclude linear combinations of inhomogeneous quantities, such as ``IP + 
$r_s$'' or ``$r_s$ + $r_p^2$''. In practice, quantities are inhomogeneous when 
they are expressed with different units. 
Except this exclusions of physically unreasonable combinations, we 
produce as many combinations as possible.
However, compressed-sensing theory poses a limit on the maximum size $M$ of the 
feature space from which the best (low-) $\Omega$-dimensional descriptor can be 
extracted by sampling the feature space with the knowledge of $N$ data points: $ 
N = C \ \! \Omega \ \! \ln (M) $ \cite{Donoho06,Candes06,Foucart06}, when the 
$M$ candidate features are uncorrelated. $C$ is not a universal constant, 
however it is typically estimated to be in the range between 4 and 8 (see 
Ref. \onlinecite{Vybiral15}). For $\Omega =2$  and $N=82$, this implies a range 
of $M$ between $\sim 200$ and $\sim 30000$. Therefore, we regarded values of 
a few thousand as an upper limit for $M$. Since the number of thinkable 
features is certainly larger than few thousands, we proceeded iteratively in 
several steps, by learning from the previous step what to put and what not in 
the candidate-feature list of the next step.
In the following, we describe how a set of $\sim 4500$ features was created. In Appendixes \ref{A:A} and \ref{A:B}, we summarize how different feature space can be constructed, starting from different assumptions. 

First of all, we form sums and absolute differences of homogeneous quantities 
and apply some unary operators (powers, exponential), see Table 
\ref{T:featspace1}.

\begin{table}[h!]
\begin{tabular}{|c|l|c|c|}
\hline
ID & Description & Prototype formula & \# \\
\hline
\parbox[c]{1.1cm} {$B1$} & Absolute differences of $A1$ & 
$|\textrm{IP}(\textrm{A})-\textrm{EA}(\textrm{A})|$ & 
\parbox[c]{1. cm} {6} \\
$B2$ & Absolute differences of $A2$ & $|\textrm{L}(\textrm{A}) - 
\textrm{H}(\textrm{A})|$ & 6\\
$B3$ & Absolute differences and sums of $A3$ & $|r_p(\textrm{A}) \pm 
r_s(\textrm{A})|$ & 30 \\
$C3$ & Squares of $A3$ and $B3$ (only sums) & $r_s(\textrm{A})^2, 
(r_p(\textrm{A}) + r_s(\textrm{A}))^2$ & 21\\
$D3$ & Exponentials of $A3$ and $B3$ (only sums) & $\exp(r_s(\textrm{A})), 
\exp(r_p(\textrm{A}) \pm r_s(\textrm{A}))$ & 21\\
$E3$ & Exponentials of squared $A3$ and $B3$ (only sums) & 
$\exp[r_s(\textrm{A})^2], \exp[(r_p(\textrm{A}) \pm r_s(\textrm{A}))^2]$ & 21\\
\hline
\end{tabular}
\caption{First set of operators applied to the primary features (Table 
\ref{T:featspace0}). Each group, labeled by a different ID, is formed by 
starting from a different group of primary features and/or by applying a 
different operator. The label A stays for both A and B of the binary material
\label{T:featspace1}}
\end{table}

Next, the above combinations are further combined, see Table \ref{T:featspace2}. 

\begin{table}[h!]
\vskip.2cm\noindent
\begin{tabular}{|c|l|c|c|}
\hline
ID & Description & Prototype formula & \# \\
\hline
$\{F1,F2,F3\}$ & Abs. differences and sums of   & $\left| |r_p(\textrm{A}) \pm 
r_s(\textrm{A})| \mp |r_p(\textrm{B}) \pm r_s(\textrm{B})| \right| $  
[\onlinecite{rsrp}] & \parbox {1.1cm} {72} \\
     & $\{B1,B2,B3\}$, without repetitions & & \\
 $G$ & ratios of any of $\{Ai,Bi\}, i=1,2,3$ & $\left| 
r_p(\textrm{B})-r_s(\textrm{B}) \right| / 
(r_d\textrm{(\textrm{A})}+r_s(\textrm{B}))^2$ & $\sim 4300$ \\
  & with any of $\{A3,C3,D3,E3\}$ & & \\
\hline
\end{tabular}
\caption{Second set of operators applied to the groups defined in Table 
\ref{T:featspace0} and \ref{T:featspace1}. 
\label{T:featspace2}}
\end{table}
LASSO was then applied to this set of $\sim 4500$ candidate features. If 
the features had low linear correlation, the first two features appearing upon 
decreasing $\lambda$ would be the best 2D descriptor, i.e., the one that 
minimizes the RMSE \cite{CorrCov}. Unfortunately, checking all pairs of features for linear correlation would scale with size $M$ as unfavorably as just performing the brute force 
search for the best 2D descriptor by trying all pairs. Furthermore, such a 
screening would require the definition of a threshold for the absolute 
value of the covariance, for the decision whether or not any two 
features are correlated, and then possibly discarding one of the two. A 
similar problem would appear in case more refined techniques, like 
singular-value decomposition, were tried in order to discard 
eigenvectors with low eigenvalues. Still a threshold should be defined and thus 
tuned.

We adopted instead a simple yet effective solution: The best $\Omega=30$ 
features with non-zero coefficients that emerge from the application of LASSO at 
decreasing $\lambda$ are grouped, and among them an exhaustive $\ell_0$ 
minimization is performed (Eqs. \eqref{eq:l0} and \eqref{eq:rr0} for 
$\|\bm{c}\|_0 = 1, 2, 3,$ \ldots). The single features, pairs, triplets, etc. 
that score the lowest RMSE are the outcome of the procedure as 1D, 2D, 3D, etc., 
descriptors. The validity of this approach was tested by checking that running 
it on smaller feature spaces ($M \sim $~few hundreds), where the direct 
search among all pairs and all triples could be carried out, gave the same 
result.

\begin{figure} [h!]
\centering
\includegraphics[width=0.85\textwidth,clip]{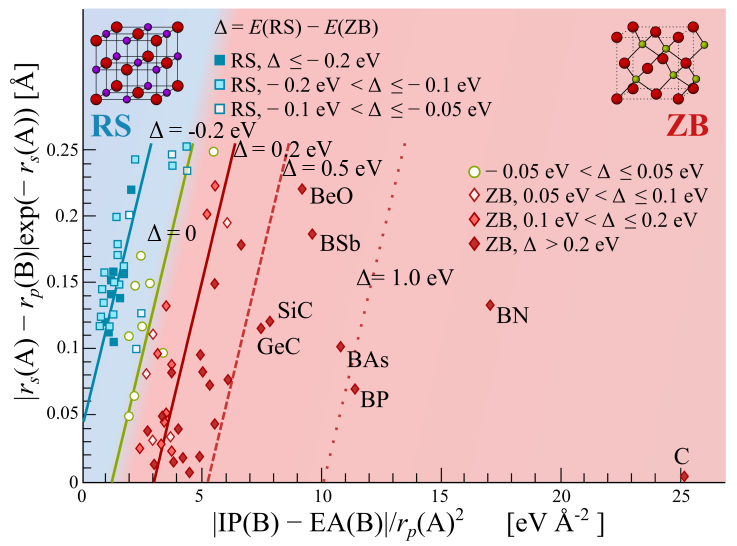}
\caption{Predicted energy differences between RS and ZB structures of the 82 
octet binary AB materials, arranged according to our optimal two-dimensional 
descriptor. The parallel straight lines are isolevels of the predicted model, 
from left to right, at -0.2, 0, 0.2, 0.5, 1.0 eV. The distance from the 0 line 
is proportional to the difference in energy between RS and ZB. The color/symbol 
code is for the reference (DFT-LDA) energies.}
\label{Fig:2D}
\end{figure}

Our procedure, applied to the above defined set of features (Tables 
\ref{T:featspace0}-\ref{T:featspace2}), found both the best 1D, 2D, and 3D 
descriptor, as well as the coefficients of the equations for predicting the 
property $\Delta E$ as shown below (energies are in eV and radii are in \AA):
\begin{eqnarray}
  \label{eq:desc1D} \Delta E &=& 0.117 
\frac{\textrm{EA(B)}-\textrm{IP(B)}}{r_p(\textrm{A})^2} - 0.342,\\
  \label{eq:desc2D} \Delta E &=& 0.113 
\frac{\textrm{EA(B)}-\textrm{IP(B)}}{r_p(\textrm{A})^2} + 1.542 
\frac{|r_s(\textrm{A})-r_p(\textrm{B})|}{\exp(r_s(\textrm{A}))} - 0.137,\\
   \nonumber \Delta E &=& 0.108 
\frac{\textrm{EA(B)}-\textrm{IP(B)}}{r_p(\textrm{A})^2} + 1.737 
\frac{|r_s(\textrm{A})-r_p(\textrm{B})|}{\exp(r_s(\textrm{A}))} + \\
  \label{eq:descr3D} & + & 9.025 \frac{\left| r_p(\textrm{B})-r_s(\textrm{B}) 
\right|}{\exp(r_d\textrm{(\textrm{A})}+r_s(\textrm{B}))} - 0.030.
\end{eqnarray}
We removed the absolute value from ``$\textrm{IP(B)}-\textrm{EA(B)}$'' as this difference is always negative.
In Fig. \ref{Fig:2D}, we show a {\em structure map} obtained by plotting the 82 
materials, where we used the two components of the 2D descriptor (Eq. 
\eqref{eq:desc2D}) as coordinates. 
We note that the descriptor we found contains physically meaningful quantities, 
like the band gap of B in the numerator of the first component 
and the size mismatch between valence $s$- and $p$-orbitals (numerators of the second and third component).

In closing this section, we note that the algorithm described above can be 
dynamically run in a web-based graphical application at:\\ 
\texttt{https://analytics-toolkit.nomad-coe.eu/tutorial-LASSO\_L0}.

\section{Cross Validation, sensitivity analysis, and extrapolation}

In this section, we discuss in detail a series of analyses performed on our 
algorithm. We start with the adopted cross-validation (CV) scheme. 
Then, we investigate how the cross-validation error varies with the size 
of the feature space (actually, its ``complexity'', as will be defined below). 
We proceed by discussing the stability of the choice of the descriptor with 
respect to sensitivity analysis. Finally, we test the {\em extrapolation} 
capabilities of the model.

In the numerical test described above, we have always used all the data for the 
training of the models; the RMSE given as figure of merit were therefore {\em 
fitting errors}. However, in order to assess the predictive ability of a model, 
it is necessary to test it on data that have not been used for the training, 
otherwise one can incur the so-called {\em overfitting}\cite{Tibshirani09}. 
Overfitting is in general signaled by a noticeable discrepancy between the 
fitting error (the RMSE over the training data) and the test error (the RMSE 
over control data, i.e., the data that were not used during the fitting 
procedure). If a large amount of data is available, one can just partition the 
data into a training and a test set, fully isolated one from another. In our 
case, that is not so unusual, the amount of data is too little for such 
partitioning, therefore, we adopted a cross-validation strategy.

In general, the data set is still partitioned into training and test data, but this procedure repeated several times, choosing different 
test data, in order to achieve a good statistics. In our case, we adopted a leave-10\%-out 
cross-validation scheme, where the data set is divided randomly into a $\sim 
90$\% of training data (75 data points, in our case) and a $\sim 10$\% of test 
data. The model is trained on the training data and the RMSE is evaluated on the 
test set. By ``training'', we mean the whole LASSO+$\ell_0$ procedure that 
selects the descriptor and determines the model (the coefficients of the linear 
equation) as in Eqs. \eqref{eq:desc1D}--\eqref{eq:descr3D}. Another figure of 
merit that was monitored is the maximum absolute error over the test set. The 
random selection, training, and error evaluation procedure was repeated 
until the average RMSE and maximum absolute errors did not change 
significantly. In practice, we typically performed 150 iterations, but the 
quantities were actually converged well before. We note that, at each iteration, 
the standardization is applied by calculating the average and 
standard deviation only of the data points in the training set. In 
this way, no information from the test set is used in the training, while if the 
standardization were computed once and for all over all the available data, some 
information on the test set would be used in the training. In fact, it can be 
shown \cite{Tibshirani09} that such approach can lead to spurious performance.

The cross-validation test can serve different purposes, depending on the adopted 
framework. For instance, in kernel ridge regression, for a Gaussian kernel, the 
fitted property is expressed as a weighted sum of Gaussians (see also section 
V): $P({\bm d}) = \sum_{i=1}^{N}  c_{i} \exp{ \left( - \| {\bm d}_{i} - {\bm 
d}\|^2_2 /  2\sigma^2 \right) } $, where $N$ is the number of training data 
points, i.e., there are as many coefficients as (training) data points.
The coefficients $c_{i}$ are determined by minimizing  $ \sum_{i=1}^{N} (P({\bm 
d}_{i}) - P_{i})^2 + \lambda \sum_{i,j=1}^{N,N} c_{i}{c}_{j} \exp{ \left( - \| 
{\bm d}_{i} - {\bm d}_{j}\|^2_2 /  2\sigma^2 \right) } $, where  $\| {\bm d}_i - 
{\bm d}_j\|^2_2 = \sum_{\alpha=1}^\Omega ( d_{i,\alpha} - d_{j,\alpha})^2$ is 
the squared $\ell_2$ norm of the difference of descriptors of different 
materials.
A recommended strategy \cite{Tibshirani09} is to use the cross validation to 
determine the optimal value of the so-called {\em hyper-parameters}, $\lambda$ 
and $\sigma$, in the sense that it is selected the pair $(\lambda,\sigma)$ that minimizes the average RMSE upon cross validation. 
In our scheme, we can 
regard the dimensionality $\Omega$ of the descriptor and the size $M$ of 
the feature space as {\em hyper-parameters}. By increasing both parameters, we 
do not observe a minimum, but we rather reach a plateau, where no significant 
improvement on the cross-validation average RMSE is achieved (see below).

Since in our procedure, the descriptor is found by the algorithm itself, a 
fundamental aspect of the cross-validation scheme is that all the procedure, 
including the selection of the descriptor, is repeated from scratch with each 
training set. This means that, potentially, the descriptor changes at each 
training-set selection. We found a remarkable stability of the 1D and 2D 
descriptors. The 1D was the same as the all-data descriptor (Eq. 
\eqref{eq:desc1D} for 90\% of the training sets, while the 2D descriptor 
was the same as in Eq. \eqref{eq:desc2D} in all cases. For 3D and higher 
dimensionality, as expected from the $ N = C \Omega \ln (M) $ relationship, the 
selection of the descriptor becomes more unstable, i.e., for different training 
sets, the selected descriptor often differs in at least one of the three 
components. The RMSE, however, does not change much from one training set to the 
other, i.e., the instability of the descriptor selection just reflects the 
presence of many competing models.
We show in Table \ref{T:tier} the cross-validation figures of merit, average 
RMSE and maximum absolute error, as a function of increased dimensionality. We 
also show in comparison the fitting error. 

\begin{table}[th!]
\centering
\begin{tabular}{|l||c|c|c|c|}
\hline
 Descriptor & \parbox[c]{1.cm} {1D} & \parbox[c]{1.cm} {2D} & \parbox[c]{1.cm} 
{3D} & \parbox[c]{1.cm} {5D} \\
\hline
RMSE & 0.14 & 0.10 & 0.08 & 0.06 \\
MaxAE & 0.32 & 0.32 & 0.24 & 0.20 \\
\hline
RMSE, CV & 0.14 & 0.11 & 0.08 & 0.07 \\
MaxAE, CV & 0.27 & 0.18 & 0.16 & 0.12 \\
\hline
\end{tabular}
\caption{Root mean square error (RMSE) and maximum absolute error (MaxAE) in eV 
for the fit of all data (first two lines) and of the test set in a 
leave-10\%-out cross validation (L-10\%-OCV), averaged over 150 random 
selections of the training set (last two lines), according to our LASSO+$\ell_0$ 
algorithm.
\label{T:RMSE}}
\end{table}

\subsection{Complexity of the feature space}
Our feature space is subdivided in 5 tiers. 
\begin{itemize}
\item 
In tier zero, we have the 14 primary features as in Table \ref{T:featspace0}. 
\item
In tier one, we group features obtained by applying only one unary (e.g., 
$\mathcal{A}^2, \exp(\mathcal{A})$) or binary (e.g., $|\mathcal{A} - 
\mathcal{B}|$) operation on primary features, where $\mathcal{A}$ and 
$\mathcal{B}$ stand for any primary feature. Note that in this scheme the 
absolute value, applied to differences, is not counted as an extra operation, 
i.e., we consider the operator $|\mathcal{A}-\mathcal{B}|$ as a single operator.
\item
In tier two, two operations are applied, e.g., 
$\mathcal{A}/(\mathcal{B}+\mathcal{C}), (\mathcal{A}-\mathcal{B})^2, 
\mathcal{A}/\mathcal{B}^2, \mathcal{A}\exp(-\mathcal{B})$. 
\item
In tier three, we apply three operations: $|\mathcal{A} \pm B 
|/(\mathcal{C}+\mathcal{D}), |\mathcal{A} \pm \mathcal{B} |/\mathcal{C}^2, 
|\mathcal{A} \pm \mathcal{B} |\exp(-\mathcal{C}), 
\mathcal{A}/(\mathcal{B}+\mathcal{C})^2, \ldots $.\\
\item
Tier four: $|\mathcal{A} \pm \mathcal{B} |/(\mathcal{C}+\mathcal{D})^2, 
|\mathcal{A} \pm \mathcal{B} |\exp(-(\mathcal{C}+\mathcal{D})), \ldots$.
\item
Tier five: $|\mathcal{A} \pm \mathcal{B} |\exp(-(\mathcal{C}+\mathcal{D})^2), 
\ldots$.
\end{itemize}
Our procedure was executed with tier 0, then with tier 0 AND 1, then 
with tiers from 0 to 2, and so on.
The results are shown in Table \ref{T:featnoise}. A clear result of this 
test is that little is gained, in terms of RMSE, when going beyond tier 3. The 
reason why MaxAE may increase at larger tiers is that the choice of the 
descriptor becomes more unstable (i.e., different descriptors may be selected) 
the larger the feature space is. This leads to less controlled 
maximal errors, and is a reflection of overfitting.

\begin{table}[h!]
\begin{tabular}{|l|r|r|r|r|r|r|}
\hline
 & Tier 0 & Tier 1 & Tier 2 & Tier 3 & Tier 4 & Tier 5 \\
 \hline
 $\Omega = 1$ & & & & & & \\
 RMSE, CV & 0.31 & 0.19 & 0.14 & 0.14 & 0.14 & 0.14 \\
 MaxAE, CV & 0.67 & 0.37 & 0.32 & 0.28 & 0.29 & 0.30 \\
\hline
 $\Omega = 2$ & & & & & & \\
 RMSE, CV & 0.27 & 0.16 & 0.12 & 0.10 & 0.10 & 0.10 \\
 MaxAE, CV & 0.60 & 0.39 & 0.27 & 0.18 & 0.19 & 0.22 \\
 \hline
 $\Omega = 3$ & & & & & & \\
 RMSE, CV & 0.27 & 0.12 & 0.10 & 0.08 & 0.08 & 0.08 \\
 MaxAE, CV & 0.52 & 0.39 & 0.27 & 0.16 & 0.18 & 0.20 \\
\hline
\end{tabular}
\caption{Errors after L-10\%-OCV. ``Tier $x$'' means that {\em all tiers up to} 
tier $x$ are included in the feature space. 
\label{T:tier}}
\end{table}

Incidentally, while for the 1D and 2D descriptor, the results presented in Eqs. 
\eqref{eq:desc1D} and \eqref{eq:desc2D} contain only features up to tier 3, the 
third component of the 3D descriptor shown in Eq. \eqref{eq:descr3D} belongs to 
tier 4. The 3D descriptor and model limited to tier 3 is:
\begin{eqnarray}
   \Delta E &=& 0.108 \frac{\textrm{EA(B)}-\textrm{IP(B)}}{r_p(\textrm{A})^2} + 
1.790 \frac{|r_s(\textrm{A})-r_p(\textrm{B})|}{\exp(r_s(\textrm{A}))} + \\
  \nonumber & + & 3.766 \frac{\left| r_p(\textrm{B})-r_s(\textrm{B}) 
\right|}{\exp(r_d\textrm{(\textrm{A})})} - 0.0267
\end{eqnarray}
This is the same as presented in Ref. \onlinecite{Ghiringhelli2015}. The CV RMSE 
of the 3D model compared to the one in Eq. \eqref{eq:descr3D} is worse by less 
than 0.01 eV.

\subsection{Sensitivity Analysis}
Cross validation tests if the found model is good only for the specific set of 
data used for the training or if it is stable enough to predict the value of the 
target property for unseen data. Sensitivity analysis is a complementary 
test on the stability of the model, where the data are perturbed, typically by 
random noise. The purpose of sensitivity analysis can be finding out 
which of the input parameters maximally affect the output of the model, but also 
how much the model depends on the specific values of the training data. In 
practice, the training data can be affected by {\it measurement errors} 
even if they are calculated by an {\em ab initio} model. This is because 
numerical approximations are used to calculate the actual values of both the 
primary features and the property. Since, through our LASSO+$\ell_0$ 
methodology, we determine functional relationships between the primary features 
and the property, applying noise to the primary features and the property is a 
way of finding out how much the found functional relationship is affected by 
numerical inaccuracies; in other words, if it is an artifact of the level of 
accuracy or a deeper, physically meaningful, relationship.

\subsubsection{Noise applied to the primary features}
In this numerical test, each of the 14 primary features of Table 
\ref{T:featspace0} was independently multiplied by Gaussian noise with mean 1 
and standard deviation $\sigma=0.001,0.01,0.03,0.05,0.1,0.13, 0.3$, respectively.
The derived features are then constructed by using these primary features 
including noise.
We also considered multiplying by Gaussian noise (same level as for the 
independent features) all features at once.
The test with independent features reflects the traditional sensitivity analysis 
test (see e.g., Ref. \onlinecite{Metiu09}), where the goal is to single out 
which input parameters maximally affect the results yielded by a model. The test 
with all features perturbed takes into account that all primary features (as 
well as the fitted property) are evaluated with the same physical model and 
computational parameters. Therefore, inaccuracies related to not fully 
converged computational settings are modeled as noise.

\begin{table}[h!]
\begin{tabular}{|l|l|r|r|r|r|r|r|r|}
\hline
\parbox{1.5cm} {Feature} & {CV scheme} & \multicolumn{7}{|c|}{Measure of Gaussian noise, $\sigma$}\\
\hline
\hline
  &  & \parbox[c]{1.1cm} {0.001}  & \parbox[c]{1.1cm}{0.010} &  
\parbox[c]{1.1cm}{0.030}  &  \parbox[c]{1.1cm}{0.050} & \parbox[c]{1.1cm}{0.100} 
&  \parbox[c]{1.1cm}{0.130} & \parbox[c]{1.1cm} {0.300} \\
 \hline
 IP(B) & LOOCV & 99 & 99 & 98 & 70 & 4 & 0 & 0 \\
 IP(B) & L-10\%-OCV  & 84 & 84 & 71 & 51 & 10 & 1 & 0 \\
 \hline
 EA(B) & LOOCV & 99 & 99 & 99 & 98 & 91 & 86 & 30 \\
 EA(B) & L-10\%-OCV  & 86 & 84 & 84 & 84 & 80 & 72 & 28 \\
 \hline
 $r_s(\textrm{A})$ & LOOCV & 99 & 99 & 99 & 99 & 96 & 61 & 0 \\
 $r_s(\textrm{A})$ & L-10\%-OCV  & 83 & 87 & 84 & 86 & 72 & 38 & 0 \\ 
 \hline
 $r_p(\textrm{A})$ & LOOCV & 99 & 98 & 86 & 64 & 2 & 0 & 0 \\
 $r_p(\textrm{A})$ & L-10\%-OCV  & 85 & 85 & 67 & 42 & 0 & 0 & 0 \\  
 \hline
 $r_p(\textrm{B})$ & LOOCV & 99 & 99 & 99 & 99 & 81 & 50 & 1 \\
 $r_p(\textrm{B})$ & L-10\%-OCV  & 86 & 85 & 86 & 83 & 72 & 53 & 2 \\
 \hline  \hline
 All 14 & LOOCV  & 99 & 98 & 70 & 11 & 0 & 0 & 0 \\
 All 14 & L-10\%-OCV  & 85 & 82 & 52 & 15 & 0 & 0 & 0 \\
 \hline
\end{tabular}
\caption{Number of times the 2D descriptor of the noiseless model (see Eq. 
\ref{eq:desc2D}) is found when noise is applied to the primary features. The noise is measured in terms of the standard deviation of the Gaussian-distributed set of 
random numbers that multiply the feature. Only 
for the 5 features contained in the 2D noiseless model, the noise affects the 
selection of the descriptor, therefore only 5 out of 14 primary features are 
listed. The last line shows the effect of noise applied to all primary features 
simultaneously. The results are displayed for both leave-one-out (LOOCV) 
and leave-10\%-out CV schemes, as indicated by the ``CV scheme'' column. 
\label{T:featnoise}}
\end{table}

Table \ref{T:best} summarizes the results. It reports the fractional number of 
times, in \%, in which the 2D descriptor of Eq. \eqref{eq:desc2D} is found by 
LASSO+$\ell_0$ as a function of the noise level. For each noise level, 50 random 
extractions of the Gaussian-distributed random number were performed. For the 
leave-one-out CV (LOOCV) scheme, 82 iterations were performed for each 
random number, i.e. each materials was once the test material. For the 
leave-10\%-out (L-10\%-OCV) 50 iterations were performed, with 50 random 
selections of 74 materials as training and 8 materials as test set. 
As expected, for the 9 primary features that do not appear in the 2D descriptor 
of Eq. \eqref{eq:desc2D}, the noise does not affect the final result at any 
level, i.e., the 2D descriptor of the noiseless model is always found, together 
with the fitting coefficients.

When one of the 5 features appearing in the 2D descriptor is perturbed, the 
result is affected by the noise level. For noise applied to some features, the 
percent of selection of the 2D descriptor of the noiseless model drops faster 
with the noise level than for others.
Of course, even when the 2D descriptor of the noiseless model is found, the 
fitting coefficients differ from iteration to iteration (each iteration is 
characterized by a different value of the random noise). For the LOOCV, the RMSE 
goes from 0.09 eV ($\sigma=0.001$) to 0.12 eV ($\sigma=0.3$), while for the 
L-10\%-OCV the RMSE goes from 0.11 to 0.15 eV. So, even when, at the largest 
noise level, the selected descriptor may vary each time, the RMSE is only mildly 
affected. This reflects the fact that many competitive models are present in the 
feature space, and therefore a model yielding a similar RMSE can always be 
found. 
It is interesting to note that upon applying noise to IP(B) in {\em all 
cases} the new 2D descriptor is 
\begin{equation}
 \left( \frac{\textrm{EA(B)}}{r_p(\textrm{A})^2}, 
\frac{|r_s(\textrm{A})-r_p(\textrm{B})|}{\exp(r_s(\textrm{A}))}
\right),
\end{equation}
i.e, the very similar to the descriptor in Eq. \eqref{eq:desc2D}, but here IP(B) is simply missing.
It is also surprising that for quite large levels of noise (10-13\%) applied to 
EA(B), the descriptor containing this feature is selected. In general, up to 
noise levels of 5\%, the descriptor of the noiseless model is recovered the 
majority of times or more. Therefore, we can conclude that the model selection 
is not very sensitive to the noise applied to isolated features. When the noise 
is applied to all features, however, the frequency of recovery of the 2D 
descriptor of Eq. \eqref{eq:desc2D} drops quickly. Still, for noise levels up to 
1\%, that could be related, e.g., to computational inaccuracies (non fully 
converged basis sets, or other numerical settings), the model is recovered 
almost always. 

\subsubsection{Adding noise to $P = \Delta E$}

We have added {\em uniformly distributed} noise of size $\delta = \pm 0.01, \pm 0.03, \ldots$ 
eV to the DFT data of $\Delta E$.
Here, we have selected two feature spaces of size 2924 and 1568, constructed by 
two different set of functions, but always including the descriptors of Eqs. 
\eqref{eq:desc1D}-\eqref{eq:descr3D}. The results are shown in Table 
\ref{T:noCBN}.
For $\Omega = 2$, we report the fraction of trials for which the 2D descriptor of the 
unperturbed data was found in a L-10\%-OCV test. (10 selection of random noise were performed and 
for each selection L-10\%-OCV was run for 50 random selections of the training set, so the statistic is over 500 independent selections.) 
The selection of the descriptor is remarkably stable up to uniform noise of $\pm 
0.1$ eV (incidentally, at around the value of the RMSE), then it drops rapidly.
We note that the errors stay constant when the noise is in the ``physically 
meaningful'' regime, i.e., the relative ordering of the materials along the 
$\Delta E$ scale is not much perturbed. Only when the noise starts mixing the 
relative order of the materials, then the prediction becomes also less and less 
accurate in terms of RMSE.

\begin{table}[h!]
\begin{tabular}{|c|l||r|r|r|r|r|}
\hline
 Number of features & Quantity & \multicolumn{5}{|c|}{ Measure of uniform noise, $\delta$ [eV]}\\
\hline
  &   & $\pm 0.0$ & $\pm 0.01$ & $\pm 0.03$ & $\pm 0.10$ & $\pm 0.20$ \\
\hline
1536 & \% best 2D descriptor & 100	& 99	& 99	& 93	& 63 \\
    & RMSE [eV] &	0.10 & 0.11 & 0.11 & 0.12 & 0.17 \\
    & AveMaxAE [eV] & 0.18 & 0.18 & 0.18 & 0.23 & 0.32 \\
\hline    
2924 & \% best 2D descriptor & 96   & 94	& 93	& 86 & 68 \\
    & RMSE [eV] &	0.11 & 0.11 & 0.12 & 0.13 & 0.18 \\
    & AveMaxAE [eV] & 0.19 & 0.20 & 0.22 & 0.24 & 0.34 \\
\hline
\end{tabular}
\caption{Performance of the model for increasing uniform noise added to the 
calculated $\Delta E$. Besides the RMSE and the AveMaxAE, we report the number of times, the 2D descriptor of the unperturbed data is recovered.
\label{T:best}}
\end{table}

\subsection{Extrapolation: (re)discovering diamond}

Most of machine-learning models, in particular kernel-based models, are known to 
yield unreliable performance for extrapolation, i.e., when predictions are made 
for a region of the input data where there are no training data. We note that, 
in condensed-matter physics, the distinction between what systems are similar 
and suitable for interpolation and what are not is difficult if not impossible. We test the 
extrapolative capabilities of our LASSO+$\ell_0$ methodology, by setting up two 
exemplary numerical tests. In the first test, we remove from the 
training set the two most stable ZB materials, namely C-diamond and BN (the two 
rightmost points in Fig. \ref{Fig:2D}), and then calculate for both of 
them $\Delta E$, as predicted by the trained model. Although the 
prediction errors of 1.2 and 0.34 eV for C and BN, respectively, are very large, 
as can be seen in Table \ref{T:noCBN} for the 2D descriptor, the model still 
predicts C and BN as the most stable ZB structures. Thus, in a setup where C 
and BN were unknown, the model would have predicted them as good candidates to 
be the most stable ZB materials.

\begin{table}[h!]
\begin{tabular}{|l||c|c||c|}
\hline
  Material & \multicolumn{2}{|c||}{$\Delta E$ [eV] } &  
\multicolumn{1}{c|}{$E_\textrm{coh}$ [eV]}\\
	\hline
	& \parbox[c]{1.7cm}{LDA}  & \parbox[c]{1.7cm}{Predicted} & 
\parbox[c]{1.7cm}{LDA}\\
  \hline
  C & -2.64 &	-1.44 & -10.14 \\
  BN & -1.71 &	-1.37 & -9.72 \\
  \hline
\end{tabular}
\caption{Performance of the model found by LASSO+$\ell_0$ when diamond and BN 
are excluded from the training. The rightmost column reports the LDA cohesive 
energy {\em per atom} of the ZB structure, referred to spinless atoms as used 
for determining the primary features in this work.
\label{T:noCBN}}
\end{table}

In the other test, we remove form the training set all four 
carbon-containing materials, namely C-diamond, SiC, GeC, and SnC, and then 
calculate for all of them $\Delta E$, as predicted by the trained model. The 
results are reported  in Table \ref{T:noCall} for the model based on the 2D 
descriptor. The prediction error for C-diamond is comparable to the first 
numerical test, and also the other errors are relatively large. However, 
the remarkable thing, here, is that the trained model does not know anything 
about carbon as a chemical element, nevertheless, it is able to predict 
that it will form ZB materials, and the relative magnitude of $\Delta E$ is 
respected.

\begin{table}[h!]
\begin{tabular}{|l||c|c||c|}
\hline
  Material & \multicolumn{2}{|c||}{$\Delta E$ [eV] } &  
\multicolumn{1}{c|}{$E_\textrm{coh}$ [eV]}\\
	\hline
	& \parbox{1.7cm}{LDA}  &  \parbox{1.7cm}{Predicted} &  
\parbox{1.7cm}{LDA}\\
  \hline
  C & -2.64 &	-1.37 & -10.14 \\
  SiC & -0.67 &	-0.48 & -8.32 \\
  GeC & -0.81 & -0.46 & -7.28 \\
  SnC & -0.45 &	-0.23 & -6.52 \\
  \hline
\end{tabular}
\caption{Performance of the model found by LASSO+$\ell_0$ when the carbon atom 
is excluded from the training, i.e., C-diamond, SiC, GeC, SnC are excluded from 
the training.The rightmost column reports the LDA cohesion energy {\em per atom} 
of the ZB structure, referred to spinless atoms as used for determining the 
primary features in this work.
\label{T:noCall}}
\end{table}

We conclude that a LASSO+$\ell_0$ based model is likely to have good, at least 
qualitative, extrapolation capabilities. This is owing to the stability 
of the linear model and the physical meaningfulness of the descriptor, which 
contains elements of the chemistry of the chemical species building the material. 

In closing this section, that we cannot draw general conclusions from 
the particularly robust performance of the descriptors that are found by 
our LASSO+$\ell_0$ algorithm when applied to the features space constructed as 
explained above. The tests described in this section, however, form a useful 
basis for assessing the robustness of a found model. We regard such or 
similar strategy to be good practice. Two criteria give us confidence that the 
found models may have a physical meaning: These are the particular nature of the 
models found by our methodology, i.e., they are expressed as explicit analytic 
functions of the primary features, and the evidence of robustness with respect 
to perturbations on the training set and the primary features. We also note 
that the functional relationships between a subset of the primary features and 
the property of interest that are found by our methodology cannot be 
automatically regarded as physical laws. In fact, both the primary features and $\Delta E$ are determined by the Kohn-Sham equations where the 
physically relevant input only consists of the atomic charges.

\section{Comparison to Gaussian-kernel ridge regression with various 
descriptors}
In this section, we use Gaussian-kernel ridge regression to predict the DFT-LDA 
$\Delta E$ for the 82 octet binaries, with various descriptors built from our 
primary features (see Table \ref{T:featspace0}) or functions of them. The 
purpose of this analysis is to point out pros and cons of using kernel ridge regression (KRR), when compared to an approach such as our 
LASSO+$\ell_0$. In the growing field of data analytics applied to 
materials-science problems, KRR is perhaps the machine-learning method 
that is most widely used to predict properties of a given set of molecules or 
materials \cite{Rupp12,Ramprasad13,Gross2014,Lilienfeld16}. 

KRR solves the nonlinear regression problem:
\begin{equation} \label{Eq:KRR}
\mathop{\rm argmin}\limits_{\bm c} \sum_{j=1}^{N} \left( P_j - \sum_{i=1}^N c_i 
k({\bm d}_i,{\bm d}_j) \right)^2 + \lambda \sum_{i,j=1}^{N} c_{i}k({\bm 
d}_i,{\bm d}_j) {c}_{j}
\end{equation}
where $P_j$ are the data points, $k({\bm d}_i,{\bm d}_j)$ is the kernel matrix 
built with the descriptor ${\bm d}$, and $\lambda$ is the regularization 
parameter, with a similar role as $\lambda$ in Eqs. \eqref{eq:rr}, 
\eqref{eq:rr0}, and \eqref{eq:rr1}. In KRR, $\lambda$ is determined by 
minimizing the cross-validation error. The fitting function determined by KRR is 
therefore $P({\bm d})= \sum_{i=1}^N c_i k({\bm d}_i,{\bm d})$, i.e. a weighted 
sum over all the data points. The crucial steps for applying this method are the 
selection of the descriptor and of the kernel. The most commonly used kernel is 
the Gaussian kernel: $k({\bm d}_i,{\bm d}_j) = \exp{ \left( - \| {\bm d}_{i} - 
{\bm d}_{j}\|^2_2 /  2\sigma^2 \right) } $. The parameter determining the width 
of the Gaussian, $\sigma$, is recommended\cite{Tibshirani09,Hansen13} to be 
determined together with $\lambda$, by minimizing the cross-validation error, 
and this is the strategy used here.
The results are summarized in Table \ref{T:KRR}. In each case, the optimal 
$(\lambda,\sigma)$ was determined by running LOOCV.

\vskip.5cm\noindent
\begin{table}[h!]
\begin{tabular}{|c|c|c|c|c|}
\hline
\parbox{1cm}{ID}  & \parbox{1cm}{Dim.} & \parbox{1.6cm}{{Descriptors}} & 
\parbox{1.cm}{$(\lambda,\sigma)$} & \parbox{1.9cm}{RMSE [eV]} \T\\
\hline
\T 1 & 2D & $Z_\textrm{A}, Z_\textrm{B}$ & $(1\cdot10^{-6},0.008)$ & 0.13 \\
2 & 2D & John and Bloch's $r_\sigma$ and $r_\pi$  & $(7\cdot10^{-6},0.008)$ & 
0.09 \\
3 & 2D & our 2D & $(7\cdot10^{-6},0.73)$ & 0.10 \\
4 & 4D & G(A), G(B), R(A), R(B) & $(1\cdot10^{-6},0.14)$ & 0.09\\
5 & 4D & $r_s(\textrm{A}), r_p(\textrm{A}), r_s(\textrm{B}), r_p(\textrm{B})$ & 
$(2\cdot10^{-6},0.14)$ & 0.08 \\
6 & 5D & IP(B), EA(B), $r_s(\textrm{A}), r_p(\textrm{A}), r_p(\textrm{B}) $ 
&$(7\cdot10^{-6},0.14)$ & 0.07\\
7 & 14D & All atomic features & $(1\cdot10^{-6},0.42)$ & 0.09 \\
8 & 23D & All atomic and dimer features & $(1\cdot10^{-6},6.72)$ & 0.24 \\
\hline
\end{tabular}
\caption{Performance of the KRR method with various descriptors, in terms of the 
minimum cross-validation RMSE over a grid of $30\times 30$ $(\lambda,\sigma)$ 
values. Descriptor 4 is built with the IUPAC group (G, 
from 1 to 18) and period (R from row, to avoid confusion with the property 
($P$)) of the elements in the PTE; performance of this set of four 
possible primary features with LASSO$+\ell_0$ is discussed in Appendix \ref{A:B1}. 
Descriptors 5 contains the 4 primary features used for building 
descriptor 2. Similarly, descriptor 6 contains the 5 primary features found in 
descriptor 3. Descriptor 7 contains all 14 atomic features listed in Table 
\ref{T:featspace0}. Descriptor 8 contains the 14 atomic features plus 9 dimer 
features (see Table \ref{T:dimers} in Appendix \ref{A:A}).
\label{T:KRR}}
\end{table}

We make the following observations:
\begin{itemize}
 \item With several atomic-based descriptors, KRR fits reach levels of RMSE 
comparable to or slightly better than our fit with the LASSO+$\ell_0$.
 \item However, the performance of KRR is not improving with the dimensionality of the descriptor: Descriptor 7 contains the same features as 
descriptor 5 or 6, plus other, possibly relevant, features. One thus expects a 
better performance, which is not the case. The same happens when going to all 23 atomic and dimer features 
(descriptor 8).  
\end{itemize}

\subsection{Prediction test with KRR}

We have repeated the tests as in section IV.C, i.e., we have 
trained a KRR model for all materials except C and BN and for all materials 
except all four carbon compounds. Then we have evaluated the predicted $\Delta 
E$ for the excluded materials.
This test was done by using descriptors 1, 2, 4, 5, and 7 from Table 
\ref{T:KRR}. Furthermore, the 2D LASSO+$\ell_0$ descriptors were evaluated for 
the two datasets described above (details on these descriptors are given in 
Appendix \ref{A:C}). These two 2D descriptors are analytical functions of five primary 
features each and these were also used as a 5D descriptor. In all cases, we have 
determined the dimensionless hyperparameters $\lambda$ and $\sigma$ by 
minimizing the RMSE over a LOOCV run, and the descriptors are normalized 
component by component. Each component of the descriptor is {\em normalized} by the $\ell_2$ norm of the vector of the values of that component 
over the whole {\em training} set. The results are shown in Tables 
\ref{T:noCBN_KRR} and \ref{T:noCx_KRR}.

\begin{table}[t!]
\begin{tabular}{|l|l|c|c|c|c|}
\hline
 Method & Descriptor &  Dim. & ($\lambda,\sigma$) & $\Delta E$(BN) & $\Delta 
E$(C) \\
 \hline
 LDA & & & & {\bf -1.71} & {\bf -2.64} \\
 \hline
 LASSO+$\ell_0$ & CBN & 2 & & -1.37 & -1.44 \\
 KRR & CBN  & 2 & $(1\cdot 10^{-6},0.14)$ & -2.85 & -4.30 \\
 KRR & CBN$^*$ & 5 & $(7\cdot 10^{-6},0.24)$ & -0.96 & -1.35 \\
 \hline
 KRR & $Z_\textrm{A}, Z_\textrm{B}$  & 2 & $(1\cdot 10^{-6},0.0085)$ & -0.68 & 
-0.56\\
 KRR & G(A), G(B), R(A), R(B) & 4 & $(1\cdot 10^{-6},0.079)$ & -0.50 & -1.29 \\
 KRR & $r_\sigma$ and $r_\pi$ & 2 & $(0.013,0.045)$ & -1.00 & -1.17 \\
 KRR & $r_s(\textrm{A}), r_p(\textrm{A}), r_s(\textrm{B}), r_p(\textrm{B})$ & 4 
& $(1\cdot 10^{-6},0.14)$ & -1.27 & -2.31 \\
 KRR & All atomic features & 14 & $(1\cdot 10^{-6},0.42)$ & -1.01 & -1.91\\
  \hline
\end{tabular}
\caption{KRR prediction of $\Delta E$ (all in eV) for C and BN, when these two 
materials are not included in the training set for various descriptors, compared 
to the LASSO$+\ell_0$ result and the LDA reference. Descriptor CBN is the same 
2D descriptor found by LASSO$+\ell_0$ for this dataset (see Table \ref{T:LMO}, row b) 
and descriptor CBN$^*$ is built with the 5 primary features found in descriptor 
CBN. 
\label{T:noCBN_KRR}}
\end{table}

\begin{table}[t!]
\begin{tabular}{|l|l|c|c|c|c|c|c|}
\hline
 Method & Descriptor &  Dim. & ($\lambda,\sigma$) & $\Delta E$(C) & $\Delta 
E$(SiC) & $\Delta E$(GeC) & $\Delta E$(SnC)\\
 \hline
 LDA & & & & {\bf -2.64} & {\bf -0.67} & {\bf -0.81} & {\bf -0.45}\\
 \hline
 LASSO+$\ell_0$ & $x$C & 2 & & -1.37 & -0.48 & -0.46 & -0.23 \\
 KRR & $x$C  & 2 & $(1.3\cdot 10^{-5},0.079)$ & -3.05 & -0.66 & -0.71 & -0.24 \\
 KRR & $x$C$^*$ & 5 & $(5.7\cdot 10^{-4},0.42)$ & -2.28 & -0.48 & -0.44 & -0.17 
\\
 \hline
 KRR & $Z_\textrm{A}, Z_\textrm{B}$  & 2 & $(7.3\cdot 10^{-3},0.015)$ & -2.38 & 
-0.22 & -0.59 & -0.29\\
 KRR & G(A),G(B),R(A),R(B) & 4 & $(1\cdot 10^{-6},0.13)$ & -2.28 & -0.48 & -0.47 
& -0.28 \\
 KRR & $r_\sigma$ and $r_\pi$ & 2 & $(1.6\cdot 10^{-4},0.079)$ & -1.96 & -0.67 & 
-0.50 & -0.31 \\
 KRR & $r_s(\textrm{A}), r_p(\textrm{A}), r_s(\textrm{B}), r_p(\textrm{B})$ & 4 
& $(1.3\cdot 10^{-5},0.079)$ & -3.06 & -0.66 & -0.70 & -0.24 \\
 KRR & All atomic features & 14 & $(1\cdot 10^{-6},0.42)$ & -1.55 & -1.24 & 
-0.31 & -0.04 \\
  \hline
\end{tabular}
\caption{KRR prediction of $\Delta E$ (all in eV) for all four carbon compounds, 
when these materials are not included in the training set for various 
descriptors, compared to the LASSO$+\ell_0$ result and the LDA reference. 
Descriptor $x$C$^*$ is the same 2D descriptor found by LASSO$+\ell_0$ for this 
dataset (see Table \ref{T:LMO}, row d) and descriptor $x$C$^*$ is built with the 5 
primary features found in descriptor $x$C.
\label{T:noCx_KRR}}
\end{table}

The performance of KRR in predicting the $\Delta E$ of selected subsets of 
materials is strongly dependent on the descriptor. In particular, when KRR is 
used for extrapolation (descriptor CBN, where C and BN data points are distant 
from the other data points in the metric defined by this 2D descriptor), the 
performance is rather poor in terms of quantitative error, even though still 
correctly predicting BN and C as very stable ZB materials. Some descriptors 
expected to carry relevant physical information, such as the set 
($r_s(\textrm{A}), r_p(\textrm{A}), r_s(\textrm{B}), r_p(\textrm{B})$), show 
also good predictive ability in these examples. 

In summary, KRR is a viable alternative to the analytical models found by 
LASSO$+\ell_0$ (as also noted in Ref. \onlinecite{Pilania2016-2}), but only when 
a good descriptor is identified. The strength of our LASSO$+\ell_0$ approach is 
that the descriptor is determined by the method itself. Most importantly, 
LASSO$+\ell_0$ is not fooled by features that are redundant or useless (i.e., 
carrying little or no information on the property). These features are simply 
discarded. In contrast, KRR cannot discard components of a descriptor, resulting 
in decreasing predictive quality when a mixture of relevant and non-relevant 
features is naively used as descriptor.

\section{Conclusions}
We have presented a compressed-sensing methodology for identifying physically 
meaningful descriptors,  i.e., physical parameters that describe the material 
and its properties of interest, and for quantitatively predicting properties relevant for materials-science. 
The methodology starts from introducing possibly relevant {\em primary features} 
that are suggested by physical intuition and pre-knowledge on the specific 
physical problem. Then, a large feature space is generated by listing nonlinear 
functions of the primary features. Finally, few features are selected with a 
compressed-sensing based method, that we call LASSO+$\ell_0$ because it uses the 
Least Absolute Shrinkage and Selection Operator for a prescreening of the 
features, and an $\ell_0$-norm minimization for the identification of the few 
most relevant features. This approach can deal well with linear correlations 
among different features. 
We analyzed the significance of the descriptors found by the LASSO+$\ell_0$ 
methodology, by discussing the interpolation ability of the model based on the 
found descriptors, the robustness of the models in terms of stability analysis, 
and their extrapolation capability, i.e., the possibility of predicting new 
materials.

\section{Acknowledgment}
This project has received funding from the European Union's Horizon 2020 
research and innovation program under grant agreement No 676580, The NOMAD 
Laboratory, a European Center of Excellence, the BBDC (contract 01IS14013E), 
and the Einstein Foundation Berlin (project ETERNAL).
J.V. was supported by the ERC CZ grant LL1203 of the Czech Ministry of Education 
and by the Neuron Fund for Support of Science. This research was initiated 
while CD, LMG, MS, and JV were visiting the Institute for Pure and Applied 
Mathematics (IPAM), which is supported by the National Science Foundation (NSF).


\begin{thebibliography}{70}
\expandafter\ifx\csname natexlab\endcsname\relax\def\natexlab#1{#1}\fi
\expandafter\ifx\csname bibnamefont\endcsname\relax
  \def\bibnamefont#1{#1}\fi
\expandafter\ifx\csname bibfnamefont\endcsname\relax
  \def\bibfnamefont#1{#1}\fi
\expandafter\ifx\csname citenamefont\endcsname\relax
  \def\citenamefont#1{#1}\fi
\expandafter\ifx\csname url\endcsname\relax
  \def\url#1{\texttt{#1}}\fi
\expandafter\ifx\csname urlprefix\endcsname\relax\def\urlprefix{URL }\fi
\providecommand{\bibinfo}[2]{#2}
\providecommand{\eprint}[2][]{\url{#2}}

\bibitem[{LB()}]{LB}
\bibinfo{note}{Springer Materials, Landolt-B\"{o}rnstein, “The most
  comprehensive collection of data in the fields of physics, physical and
  inorganic chemistry, materials science, and related fields”; http://materi-
  als.springer.com/}.

\bibitem[{\citenamefont{Lorenz et~al.}(2004)\citenamefont{Lorenz, Gro\ss{}, and
  Scheffler}}]{LorenzNN04}
\bibinfo{author}{\bibfnamefont{S.}~\bibnamefont{Lorenz}},
  \bibinfo{author}{\bibfnamefont{A.}~\bibnamefont{Gro\ss{}}}, \bibnamefont{and}
  \bibinfo{author}{\bibfnamefont{M.}~\bibnamefont{Scheffler}},
  \bibinfo{journal}{Chem. Phys. Lett.} \textbf{\bibinfo{volume}{395}},
  \bibinfo{pages}{210} (\bibinfo{year}{2004}).

\bibitem[{\citenamefont{Hautier et~al.}(2010)\citenamefont{Hautier, Fischer,
  Jain, Mueller, and Ceder}}]{Ceder10}
\bibinfo{author}{\bibfnamefont{G.}~\bibnamefont{Hautier}},
  \bibinfo{author}{\bibfnamefont{C.~C.} \bibnamefont{Fischer}},
  \bibinfo{author}{\bibfnamefont{A.}~\bibnamefont{Jain}},
  \bibinfo{author}{\bibfnamefont{T.}~\bibnamefont{Mueller}}, \bibnamefont{and}
  \bibinfo{author}{\bibfnamefont{G.}~\bibnamefont{Ceder}},
  \bibinfo{journal}{Chem. Mater.} \textbf{\bibinfo{volume}{22}},
  \bibinfo{pages}{3762} (\bibinfo{year}{2010}).

\bibitem[{\citenamefont{Behler}(2011)}]{BehelrRev11}
\bibinfo{author}{\bibfnamefont{J.}~\bibnamefont{Behler}},
  \bibinfo{journal}{Phys. Chem. Chem. Phys.} \textbf{\bibinfo{volume}{13}},
  \bibinfo{pages}{17930} (\bibinfo{year}{2011}).

\bibitem[{\citenamefont{Pilania et~al.}(2013)\citenamefont{Pilania, Wang,
  Jiang, Rajasekaran, and Ramprasad}}]{Ramprasad13}
\bibinfo{author}{\bibfnamefont{G.}~\bibnamefont{Pilania}},
  \bibinfo{author}{\bibfnamefont{C.}~\bibnamefont{Wang}},
  \bibinfo{author}{\bibfnamefont{X.}~\bibnamefont{Jiang}},
  \bibinfo{author}{\bibfnamefont{S.}~\bibnamefont{Rajasekaran}},
  \bibnamefont{and}
  \bibinfo{author}{\bibfnamefont{R.}~\bibnamefont{Ramprasad}},
  \bibinfo{journal}{Scientific Reports} \textbf{\bibinfo{volume}{3}},
  \bibinfo{pages}{2810} (\bibinfo{year}{2013}).

\bibitem[{\citenamefont{Szlachta et~al.}(2014)\citenamefont{Szlachta,
  Bart\'{o}k, and Cs\'{a}nyi}}]{Csanyi2014}
\bibinfo{author}{\bibfnamefont{W.}~\bibnamefont{Szlachta}},
  \bibinfo{author}{\bibfnamefont{A.}~\bibnamefont{Bart\'{o}k}},
  \bibnamefont{and}
  \bibinfo{author}{\bibfnamefont{G.}~\bibnamefont{Cs\'{a}nyi}},
  \bibinfo{journal}{Phys. Rev. B} \textbf{\bibinfo{volume}{90}},
  \bibinfo{pages}{104108} (\bibinfo{year}{2014}).

\bibitem[{\citenamefont{Mueller et~al.}(2014)\citenamefont{Mueller, Johlin, and
  Grossman}}]{Mueller2014}
\bibinfo{author}{\bibfnamefont{T.}~\bibnamefont{Mueller}},
  \bibinfo{author}{\bibfnamefont{E.}~\bibnamefont{Johlin}}, \bibnamefont{and}
  \bibinfo{author}{\bibfnamefont{J.}~\bibnamefont{Grossman}},
  \bibinfo{journal}{Phys. Rev. B} \textbf{\bibinfo{volume}{89}},
  \bibinfo{pages}{115202} (\bibinfo{year}{2014}).

\bibitem[{\citenamefont{Sch\"{u}tt et~al.}(2014)\citenamefont{Sch\"{u}tt,
  Glawe, Brockherde, Sanna, M\"{u}ller, and Gross}}]{Gross2014}
\bibinfo{author}{\bibfnamefont{K.}~\bibnamefont{Sch\"{u}tt}},
  \bibinfo{author}{\bibfnamefont{H.}~\bibnamefont{Glawe}},
  \bibinfo{author}{\bibfnamefont{F.}~\bibnamefont{Brockherde}},
  \bibinfo{author}{\bibfnamefont{A.}~\bibnamefont{Sanna}},
  \bibinfo{author}{\bibfnamefont{K.}~\bibnamefont{M\"{u}ller}},
  \bibnamefont{and} \bibinfo{author}{\bibfnamefont{E.}~\bibnamefont{Gross}},
  \bibinfo{journal}{Phys. Rev. B} \textbf{\bibinfo{volume}{89}},
  \bibinfo{pages}{205118} (\bibinfo{year}{2014}).

\bibitem[{\citenamefont{Atsuto et~al.}(2015)\citenamefont{Atsuto, Togo,
  Hayashi, Tsuda, Chaput, and Tanaka}}]{Tanaka2015}
\bibinfo{author}{\bibfnamefont{S.}~\bibnamefont{Atsuto}},
  \bibinfo{author}{\bibfnamefont{A.}~\bibnamefont{Togo}},
  \bibinfo{author}{\bibfnamefont{H.}~\bibnamefont{Hayashi}},
  \bibinfo{author}{\bibfnamefont{K.}~\bibnamefont{Tsuda}},
  \bibinfo{author}{\bibfnamefont{L.}~\bibnamefont{Chaput}}, \bibnamefont{and}
  \bibinfo{author}{\bibfnamefont{I.}~\bibnamefont{Tanaka}},
  \bibinfo{journal}{Phys. Rev. Lett.} \textbf{\bibinfo{volume}{115}},
  \bibinfo{pages}{205901} (\bibinfo{year}{2015}).

\bibitem[{\citenamefont{Faber et~al.}(2015)\citenamefont{Faber, Lindmaa, von
  Lilienfeld, and Armiento}}]{VonLil2015}
\bibinfo{author}{\bibfnamefont{F.}~\bibnamefont{Faber}},
  \bibinfo{author}{\bibfnamefont{A.}~\bibnamefont{Lindmaa}},
  \bibinfo{author}{\bibfnamefont{O.}~\bibnamefont{von Lilienfeld}},
  \bibnamefont{and} \bibinfo{author}{\bibfnamefont{R.}~\bibnamefont{Armiento}},
  \bibinfo{journal}{Int. J. Quantum Chemistry} \textbf{\bibinfo{volume}{115}},
  \bibinfo{pages}{1094} (\bibinfo{year}{2015}).

\bibitem[{\citenamefont{Caccin et~al.}(2015)\citenamefont{Caccin, Li, Kermode,
  and De~Vita}}]{DeVita2015}
\bibinfo{author}{\bibfnamefont{M.}~\bibnamefont{Caccin}},
  \bibinfo{author}{\bibfnamefont{Z.}~\bibnamefont{Li}},
  \bibinfo{author}{\bibfnamefont{J.}~\bibnamefont{Kermode}}, \bibnamefont{and}
  \bibinfo{author}{\bibfnamefont{A.}~\bibnamefont{De~Vita}},
  \bibinfo{journal}{Int. J. Quantum Chemistry} \textbf{\bibinfo{volume}{115}},
  \bibinfo{pages}{1129} (\bibinfo{year}{2015}).

\bibitem[{\citenamefont{Isayev et~al.}(2015)\citenamefont{Isayev, Fourches,
  Muratov, Oses, Rasch, Tropsha, and Curtarolo}}]{Curtarolo15}
\bibinfo{author}{\bibfnamefont{O.}~\bibnamefont{Isayev}},
  \bibinfo{author}{\bibfnamefont{D.}~\bibnamefont{Fourches}},
  \bibinfo{author}{\bibfnamefont{E.~N.} \bibnamefont{Muratov}},
  \bibinfo{author}{\bibfnamefont{C.}~\bibnamefont{Oses}},
  \bibinfo{author}{\bibfnamefont{K.}~\bibnamefont{Rasch}},
  \bibinfo{author}{\bibfnamefont{A.}~\bibnamefont{Tropsha}}, \bibnamefont{and}
  \bibinfo{author}{\bibfnamefont{S.}~\bibnamefont{Curtarolo}},
  \bibinfo{journal}{Chem. Mater.} \textbf{\bibinfo{volume}{27}},
  \bibinfo{pages}{753} (\bibinfo{year}{2015}).

\bibitem[{\citenamefont{Mannodi-Kanakkithodi
  et~al.}(2016)\citenamefont{Mannodi-Kanakkithodi, Pilania, Doan~Huan, Lookman,
  and Ramprasad}}]{Pilania2016-1}
\bibinfo{author}{\bibfnamefont{A.}~\bibnamefont{Mannodi-Kanakkithodi}},
  \bibinfo{author}{\bibfnamefont{G.}~\bibnamefont{Pilania}},
  \bibinfo{author}{\bibfnamefont{T.}~\bibnamefont{Doan~Huan}},
  \bibinfo{author}{\bibfnamefont{T.}~\bibnamefont{Lookman}}, \bibnamefont{and}
  \bibinfo{author}{\bibfnamefont{R.}~\bibnamefont{Ramprasad}},
  \bibinfo{journal}{Sci. Reports} \textbf{\bibinfo{volume}{6}},
  \bibinfo{pages}{20952} (\bibinfo{year}{2016}).

\bibitem[{\citenamefont{Kim et~al.}(2016)\citenamefont{Kim, Pilania, and
  Ramprasad}}]{Pilania2016-2}
\bibinfo{author}{\bibfnamefont{C.}~\bibnamefont{Kim}},
  \bibinfo{author}{\bibfnamefont{G.}~\bibnamefont{Pilania}}, \bibnamefont{and}
  \bibinfo{author}{\bibfnamefont{R.}~\bibnamefont{Ramprasad}},
  \bibinfo{journal}{Chem. Mater.} \textbf{\bibinfo{volume}{28}},
  \bibinfo{pages}{1304} (\bibinfo{year}{2016}).

\bibitem[{\citenamefont{Ueno et~al.}(2016)\citenamefont{Ueno, Rhone, Hou,
  Mizoguchi, and Tsuda}}]{COMBO2016}
\bibinfo{author}{\bibfnamefont{T.}~\bibnamefont{Ueno}},
  \bibinfo{author}{\bibfnamefont{T.}~\bibnamefont{Rhone}},
  \bibinfo{author}{\bibfnamefont{Z.}~\bibnamefont{Hou}},
  \bibinfo{author}{\bibfnamefont{T.}~\bibnamefont{Mizoguchi}},
  \bibnamefont{and} \bibinfo{author}{\bibfnamefont{K.}~\bibnamefont{Tsuda}},
  \bibinfo{journal}{Materials Discovery, in press}  (\bibinfo{year}{2016}).

\bibitem[{\citenamefont{Solovyeva and von Lilienfeld}(2016)}]{VonLil2016}
\bibinfo{author}{\bibfnamefont{A.}~\bibnamefont{Solovyeva}} \bibnamefont{and}
  \bibinfo{author}{\bibfnamefont{O.}~\bibnamefont{von Lilienfeld}},
  \bibinfo{journal}{arXiv:1605.08080}  (\bibinfo{year}{2016}).

\bibitem[{\citenamefont{Bialon et~al.}(2016)\citenamefont{Bialon,
  Hammerschmidt, and Drautz}}]{Hammerschmidt16}
\bibinfo{author}{\bibfnamefont{A.~F.} \bibnamefont{Bialon}},
  \bibinfo{author}{\bibfnamefont{T.}~\bibnamefont{Hammerschmidt}},
  \bibnamefont{and} \bibinfo{author}{\bibfnamefont{R.}~\bibnamefont{Drautz}},
  \bibinfo{journal}{Chem. Mater.} \textbf{\bibinfo{volume}{28}},
  \bibinfo{pages}{2550} (\bibinfo{year}{2016}).

\bibitem[{\citenamefont{Ghiringhelli et~al.}(2015)\citenamefont{Ghiringhelli,
  Vybiral, Levchenko, Draxl, and Scheffler}}]{Ghiringhelli2015}
\bibinfo{author}{\bibfnamefont{L.~M.} \bibnamefont{Ghiringhelli}},
  \bibinfo{author}{\bibfnamefont{J.}~\bibnamefont{Vybiral}},
  \bibinfo{author}{\bibfnamefont{S.~V.} \bibnamefont{Levchenko}},
  \bibinfo{author}{\bibfnamefont{C.}~\bibnamefont{Draxl}}, \bibnamefont{and}
  \bibinfo{author}{\bibfnamefont{M.}~\bibnamefont{Scheffler}},
  \bibinfo{journal}{Phys. Rev. Lett.} \textbf{\bibinfo{volume}{114}},
  \bibinfo{pages}{105503} (\bibinfo{year}{2015}).

\bibitem[{\citenamefont{Donoho}(2006)}]{Donoho06}
\bibinfo{author}{\bibfnamefont{D.}~\bibnamefont{Donoho}},
  \bibinfo{journal}{IEEE Trans. Inform. Theory} \textbf{\bibinfo{volume}{52}},
  \bibinfo{pages}{1289} (\bibinfo{year}{2006}).

\bibitem[{\citenamefont{Cand\`{e}s and Wakin}(2008)}]{Candes08}
\bibinfo{author}{\bibfnamefont{E.~J.} \bibnamefont{Cand\`{e}s}}
  \bibnamefont{and} \bibinfo{author}{\bibfnamefont{M.~B.} \bibnamefont{Wakin}},
  \bibinfo{journal}{IEEE Signal Processing Magazine}
  \textbf{\bibinfo{volume}{25}}, \bibinfo{pages}{21} (\bibinfo{year}{2008}).

\bibitem[{\citenamefont{Davenport et~al.}(2012)\citenamefont{Davenport, Duarte,
  Eldar, and Kutyniok}}]{Kutyniok12}
\bibinfo{author}{\bibfnamefont{M.}~\bibnamefont{Davenport}},
  \bibinfo{author}{\bibfnamefont{M.}~\bibnamefont{Duarte}},
  \bibinfo{author}{\bibfnamefont{Y.}~\bibnamefont{Eldar}}, \bibnamefont{and}
  \bibinfo{author}{\bibfnamefont{G.}~\bibnamefont{Kutyniok}},
  \emph{\bibinfo{title}{Introduction to compressed sensing: Theory and
  Applications}} (\bibinfo{publisher}{Cambridge University Press},
  \bibinfo{year}{2012}).

\bibitem[{\citenamefont{Boche et~al.}(2015)\citenamefont{Boche, Calderbank,
  Kutyniok, and Vybiral}}]{Vybiral15}
\bibinfo{author}{\bibfnamefont{H.}~\bibnamefont{Boche}},
  \bibinfo{author}{\bibfnamefont{R.}~\bibnamefont{Calderbank}},
  \bibinfo{author}{\bibfnamefont{G.}~\bibnamefont{Kutyniok}}, \bibnamefont{and}
  \bibinfo{author}{\bibfnamefont{J.}~\bibnamefont{Vybiral}},
  \emph{\bibinfo{title}{Compressed sensing and its applications}}
  (\bibinfo{year}{2015}).

\bibitem[{\citenamefont{Guyon and Elisseeff}(2003)}]{FitSel}
\bibinfo{author}{\bibfnamefont{I.}~\bibnamefont{Guyon}} \bibnamefont{and}
  \bibinfo{author}{\bibfnamefont{A.}~\bibnamefont{Elisseeff}},
  \bibinfo{journal}{J. Mach. Learn. Res.} \textbf{\bibinfo{volume}{3}},
  \bibinfo{pages}{1157} (\bibinfo{year}{2003}).

\bibitem[{\citenamefont{Kohn and Sham}(1965)}]{KohnSham65}
\bibinfo{author}{\bibfnamefont{W.}~\bibnamefont{Kohn}} \bibnamefont{and}
  \bibinfo{author}{\bibfnamefont{L.}~\bibnamefont{Sham}},
  \bibinfo{journal}{Phys. Rev.} \textbf{\bibinfo{volume}{140}},
  \bibinfo{pages}{A1133} (\bibinfo{year}{1965}).

\bibitem[{\citenamefont{Hohenberg and Kohn}(1964)}]{DFT64}
\bibinfo{author}{\bibfnamefont{P.}~\bibnamefont{Hohenberg}} \bibnamefont{and}
  \bibinfo{author}{\bibfnamefont{W.}~\bibnamefont{Kohn}},
  \bibinfo{journal}{Phys. Rev.} \textbf{\bibinfo{volume}{136}},
  \bibinfo{pages}{B864} (\bibinfo{year}{1964}).

\bibitem[{\citenamefont{Ceperley and Alder}(1980)}]{CA80}
\bibinfo{author}{\bibfnamefont{D.~M.} \bibnamefont{Ceperley}} \bibnamefont{and}
  \bibinfo{author}{\bibfnamefont{B.~J.} \bibnamefont{Alder}},
  \bibinfo{journal}{Phys. Rev. Lett.} \textbf{\bibinfo{volume}{45}},
  \bibinfo{pages}{566} (\bibinfo{year}{1980}).

\bibitem[{\citenamefont{Perdew and Wang}(1992)}]{PW92}
\bibinfo{author}{\bibfnamefont{J.~P.} \bibnamefont{Perdew}} \bibnamefont{and}
  \bibinfo{author}{\bibfnamefont{Y.}~\bibnamefont{Wang}},
  \bibinfo{journal}{Phys. Rev. B} \textbf{\bibinfo{volume}{45}},
  \bibinfo{pages}{13244} (\bibinfo{year}{1992}).

\bibitem[{\citenamefont{van Vechten}(1969)}]{vanVechten69}
\bibinfo{author}{\bibfnamefont{J.~A.} \bibnamefont{van Vechten}},
  \bibinfo{journal}{Phys. Rev.} \textbf{\bibinfo{volume}{182}},
  \bibinfo{pages}{891} (\bibinfo{year}{1969}).

\bibitem[{\citenamefont{Phillips}(1970)}]{Phillips70}
\bibinfo{author}{\bibfnamefont{J.~C.} \bibnamefont{Phillips}},
  \bibinfo{journal}{Rev. Mod. Phys.} \textbf{\bibinfo{volume}{42}},
  \bibinfo{pages}{317} (\bibinfo{year}{1970}).

\bibitem[{\citenamefont{Bloch and Simons}(1972)}]{Bloch72}
\bibinfo{author}{\bibfnamefont{A.~N.} \bibnamefont{Bloch}} \bibnamefont{and}
  \bibinfo{author}{\bibfnamefont{G.}~\bibnamefont{Simons}},
  \bibinfo{journal}{J. Am. Chem. Soc.} \textbf{\bibinfo{volume}{94}},
  \bibinfo{pages}{8611} (\bibinfo{year}{1972}).

\bibitem[{\citenamefont{John and Bloch}(1974)}]{Bloch74}
\bibinfo{author}{\bibfnamefont{J.~S.} \bibnamefont{John}} \bibnamefont{and}
  \bibinfo{author}{\bibfnamefont{A.}~\bibnamefont{Bloch}},
  \bibinfo{journal}{Phys. Rev. Lett.} \textbf{\bibinfo{volume}{33}},
  \bibinfo{pages}{1095} (\bibinfo{year}{1974}).

\bibitem[{\citenamefont{Zunger}(1980)}]{Zunger80}
\bibinfo{author}{\bibfnamefont{A.}~\bibnamefont{Zunger}},
  \bibinfo{journal}{Phys. Rev. B} \textbf{\bibinfo{volume}{22}},
  \bibinfo{pages}{5839} (\bibinfo{year}{1980}).

\bibitem[{\citenamefont{Chelikowsky}(1982)}]{Chelikowsky82}
\bibinfo{author}{\bibfnamefont{J.}~\bibnamefont{Chelikowsky}},
  \bibinfo{journal}{Phys. Rev. B} \textbf{\bibinfo{volume}{26}},
  \bibinfo{pages}{3433} (\bibinfo{year}{1982}).

\bibitem[{\citenamefont{Pettifor}(1984)}]{Pettifor84}
\bibinfo{author}{\bibfnamefont{D.~G.} \bibnamefont{Pettifor}},
  \bibinfo{journal}{Solid State Commun.} \textbf{\bibinfo{volume}{51}},
  \bibinfo{pages}{31} (\bibinfo{year}{1984}).

\bibitem[{\citenamefont{Villars}(1985)}]{Villars85}
\bibinfo{author}{\bibfnamefont{P.}~\bibnamefont{Villars}},
  \bibinfo{journal}{Less-Common Met.} \textbf{\bibinfo{volume}{109}},
  \bibinfo{pages}{93} (\bibinfo{year}{1985}).

\bibitem[{\citenamefont{Andreoni et~al.}(1985)\citenamefont{Andreoni, Galli,
  and Tosi}}]{Tosi85}
\bibinfo{author}{\bibfnamefont{W.}~\bibnamefont{Andreoni}},
  \bibinfo{author}{\bibfnamefont{G.}~\bibnamefont{Galli}}, \bibnamefont{and}
  \bibinfo{author}{\bibfnamefont{M.}~\bibnamefont{Tosi}},
  \bibinfo{journal}{Phys. Rev. Lett.} \textbf{\bibinfo{volume}{55}},
  \bibinfo{pages}{1734} (\bibinfo{year}{1985}).

\bibitem[{\citenamefont{Andreoni and Galli}(1987)}]{Galli87}
\bibinfo{author}{\bibfnamefont{W.}~\bibnamefont{Andreoni}} \bibnamefont{and}
  \bibinfo{author}{\bibfnamefont{G.}~\bibnamefont{Galli}},
  \bibinfo{journal}{Phys. Rev. Lett.} \textbf{\bibinfo{volume}{58}},
  \bibinfo{pages}{2742} (\bibinfo{year}{1987}).

\bibitem[{\citenamefont{Saad et~al.}(2012)\citenamefont{Saad, Gao, Ngo,
  Bobbitt, Chelikowsky, and Andreoni}}]{Chelikowski12}
\bibinfo{author}{\bibfnamefont{Y.}~\bibnamefont{Saad}},
  \bibinfo{author}{\bibfnamefont{D.}~\bibnamefont{Gao}},
  \bibinfo{author}{\bibfnamefont{T.}~\bibnamefont{Ngo}},
  \bibinfo{author}{\bibfnamefont{S.}~\bibnamefont{Bobbitt}},
  \bibinfo{author}{\bibfnamefont{J.~R.} \bibnamefont{Chelikowsky}},
  \bibnamefont{and} \bibinfo{author}{\bibfnamefont{W.}~\bibnamefont{Andreoni}},
  \bibinfo{journal}{Phys. Rev. B} \textbf{\bibinfo{volume}{85}},
  \bibinfo{pages}{104104} (\bibinfo{year}{2012}).

\bibitem[{\citenamefont{Bertsekas}(1999)}]{nonlinpro}
\bibinfo{author}{\bibfnamefont{D.~P.} \bibnamefont{Bertsekas}},
  \emph{\bibinfo{title}{Nonlinear Programming}} (\bibinfo{publisher}{Athena
  Scientific}, \bibinfo{year}{1999}).

\bibitem[{\citenamefont{Arora and Barak}(2009)}]{Arora09}
\bibinfo{author}{\bibfnamefont{S.}~\bibnamefont{Arora}} \bibnamefont{and}
  \bibinfo{author}{\bibfnamefont{B.}~\bibnamefont{Barak}},
  \emph{\bibinfo{title}{Computational Complexity: A Modern Approach}}
  (\bibinfo{publisher}{University Press}, \bibinfo{address}{Cambridge},
  \bibinfo{year}{2009}).

\bibitem[{\citenamefont{Tibshirani}(1996)}]{LASSO}
\bibinfo{author}{\bibfnamefont{R.}~\bibnamefont{Tibshirani}},
  \bibinfo{journal}{J. Royal Statist. Soc. B} \textbf{\bibinfo{volume}{58}},
  \bibinfo{pages}{267} (\bibinfo{year}{1996}).

\bibitem[{\citenamefont{Hastie et~al.}(2009)\citenamefont{Hastie, Tibshirani,
  and Friedman}}]{Tibshirani09}
\bibinfo{author}{\bibfnamefont{T.}~\bibnamefont{Hastie}},
  \bibinfo{author}{\bibfnamefont{R.}~\bibnamefont{Tibshirani}},
  \bibnamefont{and} \bibinfo{author}{\bibfnamefont{J.}~\bibnamefont{Friedman}},
  \emph{\bibinfo{title}{The elements of statistical learning}}
  (\bibinfo{publisher}{Springer}, \bibinfo{address}{New York},
  \bibinfo{year}{2009}).

\bibitem[{\citenamefont{Greenshtein}(2006)}]{Greenshtein06}
\bibinfo{author}{\bibfnamefont{E.}~\bibnamefont{Greenshtein}},
  \bibinfo{journal}{Ann. Statist.} \textbf{\bibinfo{volume}{34}},
  \bibinfo{pages}{2367} (\bibinfo{year}{2006}).

\bibitem[{\citenamefont{van~de Geer}(2008)}]{deGeer08}
\bibinfo{author}{\bibfnamefont{S.}~\bibnamefont{van~de Geer}},
  \bibinfo{journal}{Ann. Statist.} \textbf{\bibinfo{volume}{36}},
  \bibinfo{pages}{614} (\bibinfo{year}{2008}).

\bibitem[{\citenamefont{B\"{u}hlmann and van~de Geer}(2011)}]{deGeer11}
\bibinfo{author}{\bibfnamefont{P.}~\bibnamefont{B\"{u}hlmann}}
  \bibnamefont{and} \bibinfo{author}{\bibfnamefont{S.}~\bibnamefont{van~de
  Geer}}, \emph{\bibinfo{title}{Statistics for High-Dimensional Data}}
  (\bibinfo{publisher}{Springer}, \bibinfo{address}{Berlin},
  \bibinfo{year}{2011}).

\bibitem[{\citenamefont{Cohen et~al.}(2009)\citenamefont{Cohen, Dahmen, and
  DeVore}}]{NSP09}
\bibinfo{author}{\bibfnamefont{A.}~\bibnamefont{Cohen}},
  \bibinfo{author}{\bibfnamefont{W.}~\bibnamefont{Dahmen}}, \bibnamefont{and}
  \bibinfo{author}{\bibfnamefont{R.}~\bibnamefont{DeVore}},
  \bibinfo{journal}{J. Am. Math. Soc.} \textbf{\bibinfo{volume}{22}},
  \bibinfo{pages}{211} (\bibinfo{year}{2009}).

\bibitem[{\citenamefont{Gross et~al.}(2010)\citenamefont{Gross, Liu, Flammia,
  Becker, and Eisert}}]{Eisert10}
\bibinfo{author}{\bibfnamefont{D.}~\bibnamefont{Gross}},
  \bibinfo{author}{\bibfnamefont{Y.-K.} \bibnamefont{Liu}},
  \bibinfo{author}{\bibfnamefont{S.~T.} \bibnamefont{Flammia}},
  \bibinfo{author}{\bibfnamefont{S.}~\bibnamefont{Becker}}, \bibnamefont{and}
  \bibinfo{author}{\bibfnamefont{J.}~\bibnamefont{Eisert}},
  \bibinfo{journal}{Phys. Rev. Lett.} \textbf{\bibinfo{volume}{105}},
  \bibinfo{pages}{150401} (\bibinfo{year}{2010}).

\bibitem[{\citenamefont{Blum et~al.}(2009)\citenamefont{Blum, Gehrke, Hanke,
  Havu, Havu, Ren, Reuter, and Scheffler}}]{Blum09}
\bibinfo{author}{\bibfnamefont{V.}~\bibnamefont{Blum}},
  \bibinfo{author}{\bibfnamefont{R.}~\bibnamefont{Gehrke}},
  \bibinfo{author}{\bibfnamefont{F.}~\bibnamefont{Hanke}},
  \bibinfo{author}{\bibfnamefont{P.}~\bibnamefont{Havu}},
  \bibinfo{author}{\bibfnamefont{V.}~\bibnamefont{Havu}},
  \bibinfo{author}{\bibfnamefont{X.}~\bibnamefont{Ren}},
  \bibinfo{author}{\bibfnamefont{K.}~\bibnamefont{Reuter}}, \bibnamefont{and}
  \bibinfo{author}{\bibfnamefont{M.}~\bibnamefont{Scheffler}},
  \bibinfo{journal}{Comp. Phys. Comm.} \textbf{\bibinfo{volume}{180}},
  \bibinfo{pages}{2175} (\bibinfo{year}{2009}).

\bibitem[{\citenamefont{van Lenthe et~al.}(1994)\citenamefont{van Lenthe,
  Baerends, and Snijders}}]{scaledZORA}
\bibinfo{author}{\bibfnamefont{E.}~\bibnamefont{van Lenthe}},
  \bibinfo{author}{\bibfnamefont{E.~J.} \bibnamefont{Baerends}},
  \bibnamefont{and} \bibinfo{author}{\bibfnamefont{J.~G.}
  \bibnamefont{Snijders}}, \bibinfo{journal}{J. Chem. Phys.}
  \textbf{\bibinfo{volume}{101}}, \bibinfo{pages}{9783} (\bibinfo{year}{1994}).

\bibitem[{IPE()}]{IPEA}
\bibinfo{note}{We used for IP (EA) the energy of the half occupied Kohn-Sham
  orbital in the half positively (negatively) charged atom.}

\bibitem[{\citenamefont{Simons and Bloch}(1973)}]{Bloch73}
\bibinfo{author}{\bibfnamefont{G.}~\bibnamefont{Simons}} \bibnamefont{and}
  \bibinfo{author}{\bibfnamefont{A.~N.} \bibnamefont{Bloch}},
  \bibinfo{journal}{Phys. Rev. B} \textbf{\bibinfo{volume}{7}},
  \bibinfo{pages}{2754} (\bibinfo{year}{1973}).

\bibitem[{\citenamefont{Chelikowsky and Phillips}(1978)}]{Phillips78}
\bibinfo{author}{\bibfnamefont{J.~R.} \bibnamefont{Chelikowsky}}
  \bibnamefont{and} \bibinfo{author}{\bibfnamefont{J.~C.}
  \bibnamefont{Phillips}}, \bibinfo{journal}{Phys. Rev. B}
  \textbf{\bibinfo{volume}{17}}, \bibinfo{pages}{2453} (\bibinfo{year}{1978}).

\bibitem[{\citenamefont{Andreoni et~al.}(1979)\citenamefont{Andreoni,
  Baldereschi, Bi\'{e}mont, and Phillips}}]{Phillips79}
\bibinfo{author}{\bibfnamefont{W.}~\bibnamefont{Andreoni}},
  \bibinfo{author}{\bibfnamefont{A.}~\bibnamefont{Baldereschi}},
  \bibinfo{author}{\bibfnamefont{E.}~\bibnamefont{Bi\'{e}mont}},
  \bibnamefont{and} \bibinfo{author}{\bibfnamefont{J.~C.}
  \bibnamefont{Phillips}}, \bibinfo{journal}{Phys. Rev. B}
  \textbf{\bibinfo{volume}{20}}, \bibinfo{pages}{4814} (\bibinfo{year}{1979}).

\bibitem[{Cor({\natexlab{a}})}]{CorrBerkson}
\bibinfo{note}{Differently from H $-$ L and IP $-$ EA, which are expected, and
  are, correlated on physical grounds for all atoms, $r_s$ is not necessairly
  expected to be correlated to $r_p$. This is true for all the atoms considered
  in this work, but it could well be that such strong correlation is not valid
  for all atoms. This situation is somewhat reminiscent of the Berkson paradox,
  i.e., when a spurious correlation is found due to a biased selection of the
  data set.}

\bibitem[{\citenamefont{Hofmann et~al.}(2008)\citenamefont{Hofmann,
  Sch\"{o}lkopf, and Smola}}]{Hofmann08}
\bibinfo{author}{\bibfnamefont{T.}~\bibnamefont{Hofmann}},
  \bibinfo{author}{\bibfnamefont{B.}~\bibnamefont{Sch\"{o}lkopf}},
  \bibnamefont{and} \bibinfo{author}{\bibfnamefont{A.~J.} \bibnamefont{Smola}},
  \bibinfo{journal}{Ann. Stat.} \textbf{\bibinfo{volume}{36}},
  \bibinfo{pages}{1171} (\bibinfo{year}{2008}).

\bibitem[{tra()}]{transcorr}
\bibinfo{note}{``Having a large correlation'' is clearly not a transitive
  relationship, i.e., if the vector $\bm{a}$ is highly correlated with $\bm{b}$
  (the absolute value of the covariance is close to 1), and $\bm{b}$ is highly
  correlated to $\bm{c}$, $\bm{c}$ is not necessarily highly correlated to
  $\bm{a}$.}

\bibitem[{\citenamefont{Nelson et~al.}(2013{\natexlab{a}})\citenamefont{Nelson,
  Hart, Zhou, and Ozoli\c{n}\u{s}}}]{Hart13}
\bibinfo{author}{\bibfnamefont{L.~J.} \bibnamefont{Nelson}},
  \bibinfo{author}{\bibfnamefont{G.~L.~W.} \bibnamefont{Hart}},
  \bibinfo{author}{\bibfnamefont{F.}~\bibnamefont{Zhou}}, \bibnamefont{and}
  \bibinfo{author}{\bibfnamefont{V.}~\bibnamefont{Ozoli\c{n}\u{s}}},
  \bibinfo{journal}{Phys. Rev. B} \textbf{\bibinfo{volume}{87}},
  \bibinfo{pages}{035125} (\bibinfo{year}{2013}{\natexlab{a}}).

\bibitem[{\citenamefont{Nelson et~al.}(2013{\natexlab{b}})\citenamefont{Nelson,
  Ozoli\c{n}\v{s}, Reese, Zhou, and Hart}}]{Ozolins13}
\bibinfo{author}{\bibfnamefont{L.}~\bibnamefont{Nelson}},
  \bibinfo{author}{\bibfnamefont{V.}~\bibnamefont{Ozoli\c{n}\v{s}}},
  \bibinfo{author}{\bibfnamefont{C.}~\bibnamefont{Reese}},
  \bibinfo{author}{\bibfnamefont{F.}~\bibnamefont{Zhou}}, \bibnamefont{and}
  \bibinfo{author}{\bibfnamefont{G.~L.~W.} \bibnamefont{Hart}},
  \bibinfo{journal}{Phys. Rev. B} \textbf{\bibinfo{volume}{88}},
  \bibinfo{pages}{155105} (\bibinfo{year}{2013}{\natexlab{b}}).

\bibitem[{\citenamefont{Ozoli\c{n}\v{s}
  et~al.}(2013)\citenamefont{Ozoli\c{n}\v{s}, Lai, Caflisch, and
  Osher}}]{OzolinsPNAS13}
\bibinfo{author}{\bibfnamefont{V.}~\bibnamefont{Ozoli\c{n}\v{s}}},
  \bibinfo{author}{\bibfnamefont{R.}~\bibnamefont{Lai}},
  \bibinfo{author}{\bibfnamefont{R.}~\bibnamefont{Caflisch}}, \bibnamefont{and}
  \bibinfo{author}{\bibfnamefont{S.}~\bibnamefont{Osher}},
  \bibinfo{journal}{Proc. Natl. Acad. Sci. USA} \textbf{\bibinfo{volume}{110}},
  \bibinfo{pages}{18368} (\bibinfo{year}{2013}).

\bibitem[{\citenamefont{Zhou et~al.}(2014)\citenamefont{Zhou, Nielson, Xia, and
  Ozoli\c{n}\v{s}}}]{Ozolins14}
\bibinfo{author}{\bibfnamefont{F.}~\bibnamefont{Zhou}},
  \bibinfo{author}{\bibfnamefont{W.}~\bibnamefont{Nielson}},
  \bibinfo{author}{\bibfnamefont{Y.}~\bibnamefont{Xia}}, \bibnamefont{and}
  \bibinfo{author}{\bibfnamefont{V.}~\bibnamefont{Ozoli\c{n}\v{s}}},
  \bibinfo{journal}{Phys. Rev. Lett.} \textbf{\bibinfo{volume}{113}},
  \bibinfo{pages}{185501} (\bibinfo{year}{2014}).

\bibitem[{\citenamefont{Budich et~al.}(2014)\citenamefont{Budich, Eisert,
  Bergholtz, Diehl, and Zoller}}]{Eisert14}
\bibinfo{author}{\bibfnamefont{J.}~\bibnamefont{Budich}},
  \bibinfo{author}{\bibfnamefont{J.}~\bibnamefont{Eisert}},
  \bibinfo{author}{\bibfnamefont{E.}~\bibnamefont{Bergholtz}},
  \bibinfo{author}{\bibfnamefont{S.}~\bibnamefont{Diehl}}, \bibnamefont{and}
  \bibinfo{author}{\bibfnamefont{P.}~\bibnamefont{Zoller}},
  \bibinfo{journal}{Phys. Rev. B} \textbf{\bibinfo{volume}{11}},
  \bibinfo{pages}{115110} (\bibinfo{year}{2014}).

\bibitem[{\citenamefont{J.R.}(1998)}]{Koza98}
\bibinfo{author}{\bibfnamefont{K.}~\bibnamefont{J.R.}},
  \emph{\bibinfo{title}{Genetic Programming}} (\bibinfo{publisher}{MIT PRess},
  \bibinfo{year}{1998}).

\bibitem[{\citenamefont{Cand\'es et~al.}(2006)\citenamefont{Cand\'es, Romberg,
  and Tao}}]{Candes06}
\bibinfo{author}{\bibfnamefont{E.~J.} \bibnamefont{Cand\'es}},
  \bibinfo{author}{\bibfnamefont{J.}~\bibnamefont{Romberg}}, \bibnamefont{and}
  \bibinfo{author}{\bibfnamefont{T.}~\bibnamefont{Tao}}, \bibinfo{journal}{IEEE
  Trans. Inform. Theory} \textbf{\bibinfo{volume}{52}}, \bibinfo{pages}{489}
  (\bibinfo{year}{2006}).

\bibitem[{\citenamefont{Foucart and Rauhut}(2013)}]{Foucart06}
\bibinfo{author}{\bibfnamefont{S.}~\bibnamefont{Foucart}} \bibnamefont{and}
  \bibinfo{author}{\bibfnamefont{H.}~\bibnamefont{Rauhut}},
  \emph{\bibinfo{title}{A mathematical introduction to compressive sensing}}
  (\bibinfo{publisher}{Springer}, \bibinfo{address}{New York},
  \bibinfo{year}{2013}).

\bibitem[{rsr()}]{rsrp}
\bibinfo{note}{With signs $(-,+,-)$ this is St. John \& Bloch's $r_\pi$, with
  signs $(+,-,+)$ it is $r_\sigma$}.

\bibitem[{Cor({\natexlab{b}})}]{CorrCov}
\bibinfo{note}{Correlation between pairs of features means in this context that
  the absolute value of the covariance (i.e., the scalar product of the
  82-dimensional vectors containing the values of features $i$ and $j$, with
  each vector subtracted of its mean value and divided by its standard
  deviation) is close to 1. If their covariance is close to zero, the two
  features are uncorrelated.}

\bibitem[{\citenamefont{Meskine et~al.}(2009)\citenamefont{Meskine, Matera,
  Scheffler, Reuter, and Metiu}}]{Metiu09}
\bibinfo{author}{\bibfnamefont{H.}~\bibnamefont{Meskine}},
  \bibinfo{author}{\bibfnamefont{S.}~\bibnamefont{Matera}},
  \bibinfo{author}{\bibfnamefont{M.}~\bibnamefont{Scheffler}},
  \bibinfo{author}{\bibfnamefont{K.}~\bibnamefont{Reuter}}, \bibnamefont{and}
  \bibinfo{author}{\bibfnamefont{H.}~\bibnamefont{Metiu}},
  \bibinfo{journal}{Surf. Sci.} \textbf{\bibinfo{volume}{603}},
  \bibinfo{pages}{1724} (\bibinfo{year}{2009}).

\bibitem[{\citenamefont{Rupp et~al.}(2012)\citenamefont{Rupp, Tkatchenko,
  M\"{u}ller, and von Lilienfeld}}]{Rupp12}
\bibinfo{author}{\bibfnamefont{M.}~\bibnamefont{Rupp}},
  \bibinfo{author}{\bibfnamefont{A.}~\bibnamefont{Tkatchenko}},
  \bibinfo{author}{\bibfnamefont{K.-R.} \bibnamefont{M\"{u}ller}},
  \bibnamefont{and} \bibinfo{author}{\bibfnamefont{O.~A.} \bibnamefont{von
  Lilienfeld}}, \bibinfo{journal}{Phys. Rev. Lett.}
  \textbf{\bibinfo{volume}{108}}, \bibinfo{pages}{058301}
  (\bibinfo{year}{2012}).

\bibitem[{\citenamefont{Faber et~al.}(2016)\citenamefont{Faber, Lindmaa, von
  Lilienfeld, and Armiento}}]{Lilienfeld16}
\bibinfo{author}{\bibfnamefont{F.~A.} \bibnamefont{Faber}},
  \bibinfo{author}{\bibfnamefont{A.}~\bibnamefont{Lindmaa}},
  \bibinfo{author}{\bibfnamefont{O.~A.} \bibnamefont{von Lilienfeld}},
  \bibnamefont{and} \bibinfo{author}{\bibfnamefont{R.}~\bibnamefont{Armiento}},
  \bibinfo{journal}{Phys. Rev. Lett.} \textbf{\bibinfo{volume}{117}},
  \bibinfo{pages}{135502} (\bibinfo{year}{2016}).

\bibitem[{\citenamefont{Hansen et~al.}(2013)\citenamefont{Hansen, Montavon,
  Biegler, Fazli, Rupp, Scheffler, von Lilienfeld, Tkatchenko, and
  M\"{u}ller}}]{Hansen13}
\bibinfo{author}{\bibfnamefont{K.}~\bibnamefont{Hansen}},
  \bibinfo{author}{\bibfnamefont{G.}~\bibnamefont{Montavon}},
  \bibinfo{author}{\bibfnamefont{F.}~\bibnamefont{Biegler}},
  \bibinfo{author}{\bibfnamefont{S.}~\bibnamefont{Fazli}},
  \bibinfo{author}{\bibfnamefont{M.}~\bibnamefont{Rupp}},
  \bibinfo{author}{\bibfnamefont{M.}~\bibnamefont{Scheffler}},
  \bibinfo{author}{\bibfnamefont{O.~A.} \bibnamefont{von Lilienfeld}},
  \bibinfo{author}{\bibfnamefont{A.}~\bibnamefont{Tkatchenko}},
  \bibnamefont{and} \bibinfo{author}{\bibfnamefont{K.-R.}
  \bibnamefont{M\"{u}ller}}, \bibinfo{journal}{JCTC}
  \textbf{\bibinfo{volume}{9}}, \bibinfo{pages}{3404} (\bibinfo{year}{2013}).

\end{thebibliography}

\appendix
\section{More details on the construction of the feature space}
\label{A:A}

In order to determine the final feature space as described in section III, we 
proceeded in this way:
\begin{itemize}
  \item As scalar features describing the valence orbitals, we use the radii at 
which the radial probability densities of the valence $s$, $p$, and $d$ orbitals 
have their maxima. This type of radii was, in fact, selected by our procedure, as 
opposed to the average radii (i.e., the quantum-mechanical expectation value of 
the radius). To the purpose, first two feature spaces starting from both sets of 
radii as primary features were constructed. In practice, in one case we started 
from the same primary features as in Table \ref{T:featspace0}, but without group 
A2 (in order to reduce the dimensionality of the final space), in the other case 
we substituted the average radii in group $A3$, again without group $A2$. We then 
constructed both spaces following the rules of Tables \ref{T:featspace1} and \ref{T:featspace2}. 
Finally, we joined the spaces and applied LASSO+$\ell_0$ to this joint space. Only features 
containing the radii at maximum were selected among the best.
  \item Similarly, we also defined three other radii-derived features for the 
atoms: the radius of the highest occupied orbital of the neutral atom, $r_0$, 
and analogously defined radii for the anions, $r_-$, and the cations, 
$r_+$. $r_0$ is either $r_s$ or $r_p$ as defined above, depending on which valence orbital is the HOMO. 
As in the previous point, we constructed a feature space containing both 
$\{r_0, r_-, r_+\}$ and $\{r_s, r_p, r_d\}$ and their combinations, and found 
that only the latter radii were selected among the best.
  \item We have considered in addition features related to the AA, BB and AB 
dimers (see Table \ref{T:dimers}). These new features where combined in the same way 
as the groups $A1$, $A2$, and $A3$, respectively (see Tables \ref{T:featspace0}--\ref{T:featspace2}). 
After running our procedure, we found that features containing dimer-related quantities are never 
selected among the most prominent ones.
  
\begin{table}[h!]
\begin{tabular}{|c|l|l|c|}
\hline
\parbox{1cm} {ID} & {Descriptor} & Symbols & \parbox{1cm}{\#} \T\\
\hline
$A1'$ & Binding energy & $E_b(\textrm{AA})$, $E_b(\textrm{BB})$,
$E_b(\textrm{AB})$ & 3 \T \\
$A2'$ & HOMO-LUMO KS gap & HL(AA), HL(BB), HL(AB) & 3 \T \\
$A3'$ & Equilibrium distance & $d$(AA), $d$(BB), $d$(AB) & 3 \T \\
 \hline
\end{tabular}
\caption{Set of primary features related to homo- and hetero-nuclear (spinless) dimers.
\label{T:dimers}}
\end{table}
\vskip.3cm

  \item We constructed in sequence several alternative sets of features, in particular 
varying systematically the elements of group $G$ (see Table \ref{T:featspace2}). 
Multiplication of the $\{Ai,Bi,Ei\}$, $(i=1,2,3)$, by the $\{A3,B3\}$ was 
included, as well as division of $\{Ai,Bi,Ei\}, (i=1,2,3)$, by the $\{A3,B3\}$ 
cubed (instead of squared ans in Table \ref{T:featspace2}). Only division by $C3$ were selected by LASSO.
  At this stage, a descriptor in the form
  \begin{equation}
 	\left(
  \frac{|\textrm{IP(B)}-\textrm{EA(B)}|}{r_p(\textrm{A})^2},\, 
\frac{|r_s(\textrm{A})-r_p(\textrm{B})|}{r_s(\textrm{A})^2},\, \frac{\left| 
r_p(\textrm{B})-r_s(\textrm{B}) \right|}{r_d\textrm{(\textrm{A})}^2}
  \right)
	\end{equation}
was found.
	
 The persistence of the $C3$ group in the denominator suggested to try other 
decaying functions of $r$ and $r+r'$; for instance, exponentials as defined in 
$D3$ and $E3$. Interestingly, when the set of features containing $C3$, $D3$, 
and $E3$ was searched, the second and third component of the above descriptor 
were substituted by corresponding forms where the denominator squared is 
replaced by exponentials of the same atomic features (see below). This 
descriptor was therefore found by LASSO, in the sense that the substitutions 
$1/r_s(\textrm{A})^2 \rightarrow \exp(-r_s(\textrm{A}))$ and 
$1/r_d\textrm{(\textrm{A})}^2 \rightarrow \exp(-r_d\textrm{(\textrm{A})})$ are 
an outcome of the LASSO procedure, not of a directly attempted substitution.
 \item The fact that all selected features belong to group $G$ (see Table 
\ref{T:featspace2}), which is the most populated, is {\em not} due to the fact 
that members of the other groups are ``submerged'' by the large population of 
group $G$ and ``not seen'' by LASSO. We have run extensive tests on the groups 
excluding $G$ and, indeed, the best-performing descriptors yield RMSE larger than 
those we have found.
 \item By noticing that the features in group $A2$ (see Table 
\ref{T:featspace0}) never appear in the selected descriptors, and that the 
information carried by the features in $A1$ is similar to those in $A2$, we 
investigated what happens if $A1$ is removed from the primary features (and 
therefore all derived features containing primary features from $A1$ are removed 
from the feature space). We find the following models
 \begin{eqnarray}
  \label{eq:desc1Db} \Delta E &=& 0.518  \frac{| \textrm{H(A)}- 
\textrm{L(B)}|}{\exp[r_p(\textrm{A})^2]} - 0.174, \\ 
  \label{eq:desc2Db} \Delta E &=& 4.335 
\frac{r_d(\textrm{A})-r_s(\textrm{B})}{(r_p(\textrm{A})+r_d(\textrm{A}))^2} + 
16.321 \frac{r_s(\textrm{B})}{\exp[(r_p(\textrm{A}) + r_p(\textrm{B}))^2]} - 
0.406.
\end{eqnarray}

In practice, $| \textrm{H(A)}- \textrm{L(B)}|$ replaces 
$|\textrm{IP(B)}-\textrm{EA(B)}|$ in the 1D model, with a similar denominator as 
in Eq. \eqref{eq:desc2D}. However, the 2D model does not contain any feature 
from $A2$ and it is all built with features from $A3$. The RMSE of the 1D model 
of Eq. \eqref{eq:desc1Db} is 0.145 eV, slightly worse than 0.142 eV for the 
model of Eq. \eqref{eq:desc1D}. The RMSE of the 2D model in Eq. 
\eqref{eq:desc2Db} is 0.109 eV, compared to 0.099 eV for Eq. \eqref{eq:desc2D}. 
Equation \eqref{eq:desc2Db} is also the best LASSO+$\ell_0$ model if 
the features space is constructed by using only $A3$ as primary features. In the 
latter case, the 1D model would be (RMSE=0.160 eV):
\begin{eqnarray}
  \label{eq:desc1Dc} \Delta E &=& 19.384 
\frac{r_p(\textrm{B})}{\exp[(r_s(\textrm{A})+r_s(\textrm{B}))^2]} - 0.257
\end{eqnarray} 
\end{itemize}

\section{Other feature spaces}
\label{A:B}

In this Appendix, we show the performance of our algorithm with feature spaces 
constructed from completely different {\em primary features} than in the main 
text. The purpose is to underline the importance of the choice of the initial 
set of features. We note that when combination rules are established, performing 
new numerical tests takes just the (human) time to tabulate the new set of 
primary features for the data set. 

\subsection{Primary features including valence and row of the PTE}
\label{A:B1}

We included the ``periodic-table coordinates'', period (labeled as R for Row) and group 
(labeled as G, from 1 to 18 according to the IUPAC numbering) as features.
The reason for this test is to see whether by introducing more information than 
just the atomic number Z$_\textrm{A}$ and Z$_\textrm{B}$ (see 
\onlinecite{Ghiringhelli2015}), a predictive model for $\Delta E$ is obtained. The new information 
is the explicit similarity between elements of the same group (they share coordinate G), 
that is not contained in the atomic number Z. 
First we started with only R(A), R(B), G(A), and G(B) as primary features and 
constructed a features space using the usual combination rules. We note that 
G(B) is redundant for many but not all the cases, given G(A). In fact, for $sp$ 
octet binaries, the sum of the groups is 18, but for binaries containing Cu, Ag, 
Cd, and Zn (16 out of 82), the sum is different and therefore the coordinates 
are effectively 4.
Next, we construct a feature space starting from the set of 14 primary fetures described in Table 
\ref{T:featspace0} -- but where the group $A2$ (the HOMO and LUMO Kohn-Sham levels) are substituted with 
R(A), R(B), G(A), and G(B) -- and then following the rules 
summarized in Tables \ref{T:featspace1}and \ref{T:featspace2}.

For the case where the primary features are only R(A), R(B), G(A), and G(B), we generated a feature space of size 
1143 and then ran LASSO+$\ell_0$. The outcome is shown in Table \ref{T:PeriodicTableFeat}. 
We conclude that R(A), R(B), G(A), and G(B) alone do not contain enough 
information for building a predictive model, following our algorithm.

\begin{table}[h!]
\begin{tabular}{|r|r|r|r|}
\hline
  & $\Omega=1$ & $\Omega=2$ & $\Omega=3$   \\
  \hline
 RMSE [eV]&  0.20 & 0.19 & 0.17 \\
MaxAE [eV]&  0.69 & 0.71 & 0.62 \\
 \hline
\end{tabular}
\caption{Feature space with primary features R (Row or Period) and G (Group) of the PTE. Performance over the whole set of 82 materials. 
\label{T:PeriodicTableFeat}}
\end{table}

For the case where the primary features are 14, including R(A), R(B), G(A), and G(B), we constructed a feature space of size 4605, and LASSO+$\ell_0$ found the following 
1D optimal model:
\begin{equation}
 \Delta E = -0.376 + 0.0944 
\frac{|\textrm{R(B)}-\textrm{G(B)}|}{r_p(\textrm{A})^2}.
\end{equation}
In essence, $|\textrm{R(B)}-\textrm{G(B)}|$ replaced 
$|\textrm{IP(B)}-\textrm{EA(B)}|$ in Eq. \eqref{eq:desc1D}, while the 
denominator remained unchanged. The RMSE of this new 1D model is only sightly 
better than the model in Eq. \eqref{eq:desc1D}, namely 0.13 eV (vs 0.14 eV), but 
the MaxAE is much worse, i.e., 0.43 vs. 0.32 eV.
However, this finding is remarkable as it implies some correlation between 
$|\textrm{R(B)}-\textrm{G(B)}|$ and $|\textrm{IP(B)}-\textrm{EA(B)}|$, where the 
latter is a DFT-LDA property while the former comes simply from the number of 
electrons and the Aufbau principle.
Indeed, there is a Pearson's correlation of 0.87 between the two quantities, 
while, when both quantities are divided by $r_p(\textrm{A})^2$, the Pearson's 
correlation becomes as high as 0.98.
The 2D and 3D optimal descriptor, though, are the same as in Eqs. 
\eqref{eq:desc2D} and \eqref{eq:descr3D}.

\begin{figure} [h!]
\centering
\includegraphics[width=0.8\textwidth,clip]{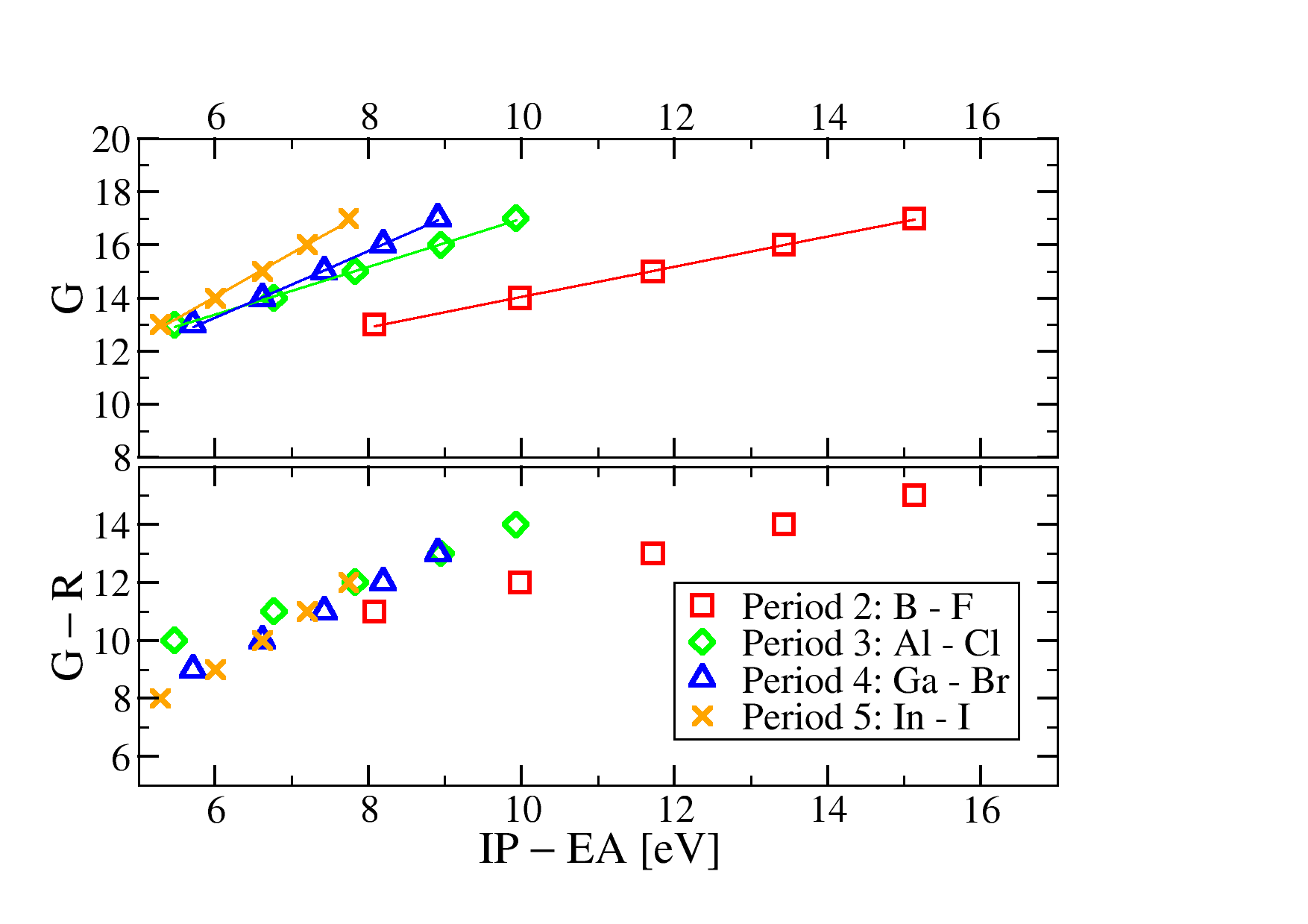}
\caption{Plot of the difference IP$-$EA for the anions (B atoms in the AB materials) vs their Group (top pane) or the difference Group$-$Row (bottom pane) in the PTE. The straight lines are linear least-square fit of the data points \label{Fig:EAIP}}
\end{figure}

Even though it is expected that the difference IP$-$EA grows when moving 
in the PTE to the right, a linear correlation between the difference $|\textrm{R(B)}-\textrm{G(B)}|$ and IP$-$EA it is unexpected.
Figure \ref{Fig:EAIP} shows in the bottom panel this correlation for the $p$ elements (the anions B in the AB materials) of the PTE, while the top panel shows an even stronger 
linear correlation between IP$-$EA and the group G in the PTE, for each period.

\subsection{Adding John and Bloch's $r_\sigma$ and $r_\pi$ as primary features}
\label{A:B2}

Here, we added $r_\sigma$ and $r_\pi$ \cite{Bloch74} to the primary features, in 
order to see whether combined with other features, according to our rules, yielded a more accurate and/or more robust modeling. The feature space was reduced, 
as plainly adding all the combinations with these 
2 extra features, made the whole procedure unstable (remember that we have only 82 data points).
By optimizing over a feature space of size 2924 and using all 82 materials for learning and testing, LASSO$+\ell_0$ identified the same 2D descriptor as in Eq. \eqref{eq:desc2D}. In other words, no function of $r_\sigma$ or $r_\pi$ won over the known descriptor. For the L-10\%-OCV, in about 10\% of the cases, a descriptor containing a 
function of $r_\sigma$ or $r_\pi$ was selected.

\subsection{Use of force constants derived from the tedrahedral pentamers AB$_4$ 
and BA$_4$ as primary features}
\label{A:B3}

Here, we build a feature space exploring the idea that the difference in energy 
between RS and ZB may depend on the mechanical stability of the basic 
constituent of either crystal structure. For instance, we choose ZB and we look 
at the mechanical stability of the tetrahedral pentamers AB$_4$ and BA$_4$. In 
practice, we look at the elastic constants for the symmetric and antisymmetric 
stretching deformations. 

We write the elastic energy of deformation of a tetrahedral pentamer XY$_4$, for 
a symmetric stretch:
\begin{equation} \label{Eq:DES}
\Delta E_\textrm{harm}^\textrm{SYMM} = 4 \alpha_\textrm{XY} (\Delta r)^2
\end{equation}
where $\alpha_\textrm{XY}$ is the bond-stretching constant (of one XY bond  
in the tetrahedral arrangement, which is not necessarily the same value as the 
one of the XY dimer). The factor 4 comes from the fact that there are 4 bonds equally stretched.
The second derivative of $\Delta E_\textrm{harm}^\textrm{SYMM}$ with respect to 
$\Delta r$ is $(\Delta E_\textrm{harm}^\textrm{SYMM})'' = 8 \alpha_\textrm{XY}$. 
The left-hand side is evaluated from the (LDA) calculated $\Delta 
E_\textrm{harm}^\textrm{SYMM}(\Delta r)$ curve, fitted to a second order 
polynomial.

Considering an asymmetric stretch, which means that we moved the central 
atom X along one of the four XY bonds, we write:
\begin{eqnarray}
 \Delta E_\textrm{harm}^\textrm{ASYMM} &=& 2 \alpha_\textrm{XY} (\Delta r)^2 + 3 
\beta_\textrm{YXY} (\theta_1)^2 + 3 \beta_\textrm{YXY} (\theta_2)^2, 
\end{eqnarray}
where $\theta_1$ and $\theta_2$ are the two different distortion angles, that 
the 6 angles formed by the atoms X-Y-X, undergo upon the asymmetric stretch. 
After working out the geometrical relationship $\theta_{1,2}^2= \theta_{1,2}^2((\Delta r)^2)$, we 
find:
\begin{eqnarray}
 \Delta E_\textrm{harm}^\textrm{ASYMM}  &=& 2 ( \alpha_\textrm{XY} + 4 
\beta_\textrm{YXY} ) (\Delta r)^2.
\end{eqnarray}
Since the asymmetric deformation is defined through $\Delta r$, for convenience, 
$\beta_\textrm{YXY}$ is referred to the linear displacement, rather than the 
angular one. In practice, the relationship $\theta^2((\Delta r)^2)$ was derived. 
The second derivative of the above expression is equated to the second 
derivative of the calculated $\Delta E (\Delta r)$ curve. Since 
$\alpha_\textrm{XY}$ is known from Eq. \eqref{Eq:DES}, $\beta_\textrm{YXY}$ is 
then inferred.

We have 4 primary features for each material, now. The first two,
$\alpha_\textrm{AB}$ and $\alpha_\textrm{ABA}$, come from the AB$_4$ tetrahedral 
molecule, while the third and fourth ones, $\alpha_\textrm{BA}$ and $\alpha_\textrm{BAB}$ come 
from the BA$_4$ molecule. In addition, by noting that $\alpha_\textrm{XY}$ and 
$\beta_\textrm{YXY}$ enter the expression $ \Delta E_\textrm{harm}^\textrm{ASYMM}$, with a ratio of 1/4, 
two primary features were added, i.e., $\alpha_\textrm{AB}+4\beta_\textrm{ABA}$ and 
$\alpha_\textrm{BA}+4\beta_\textrm{BAB}$. These 6 primary features were combined in (some of) the usual ways.
Note that the above combination of $\alpha_\textrm{BA}$ and 
$\alpha_\textrm{BAB}$ opens a new level of complexity. At present, when 
constructing the feature space, we apply operations like $\textrm{A}+\textrm{B}$ 
and $|\textrm{A}-\textrm{B}|$, but we do not allow for the freedom of arbitrary linear 
combinations of them. 
Note also that the information about the two atoms here are intermingled in all 
six primary features.

To give an idea, the so obtained 2D descriptor is:
\begin{equation}
 \left( \left[ \beta_\textrm{ABA}-(\alpha_\textrm{BA}+4\beta_\textrm{BAB}) \right]
\alpha_\textrm{AB}, 
\frac{\alpha_\textrm{AB}+4\beta_\textrm{ABA}+\alpha_\textrm{BA}+4\beta_\textrm{
BAB}}{\alpha_\textrm{AB}+\alpha_\textrm{BA}} \right)
\end{equation}
The performance in terms of RMSE and CV is summarized in Table \ref{T:elastic}.
\begin{table}[h!]
\begin{tabular}{|c|c|c|c|c|}
\hline
 \parbox{1cm}{$\Omega$} &	RMSE all &	RMSE CV &	MAE CV &	MaxAE CV \\
\hline
 $1$ &	0.1910 &	0.1809 &	0.1473 &	0.3583 \\
 $2$ &	0.1532 &	0.1797 &	0.1303 &	0.3836 \\
 $3$ &	0.1346 &	0.2195 &	0.1553 &	0.4810 \\
\hline
\end{tabular}
\caption{Performance of the model built from primary features based on force-constants. \label{T:elastic}}
\end{table}
We conclude that the features based on the elastic energy do not yield a model with good predictive ability.

\section{Numerical tests excluding selected elements}
\label{A:C}

In this numerical test, we have removed several sets of materials from the data 
set. In practice, we have removed: 

\noindent
0) Nothing \\
a) C-diamond (1 material) \\
b) C-diamond and BN (2 materials) \\
c) BN (1 material) \\
d) All carbon compounds (4 materials) \\
e) All boron compounds (4 materials) \\
f) CdO (example of a material with $\Delta E \sim 0 $ (1 material)) \\
g) All oxygen-containing compounds (7 materials) \\
h) All cadmium-containing compounds (4 materials) \\
After the removal, the usual L-10\%-OCV was performed on the remaining materials.
The purpose was to analyze the stability of the model when some crucial (or less 
crucial) elements/materials are removed completely from the data set.
Also, we aimed of identifying outliers, i.e. data points whose removal from the 
set improve the accuracy of the fit. The results for a tier-3 feature 
space of size 2420 are summarized in Table \ref{T:LMO}.
\begin{table}[h!]
\renewcommand{\arraystretch}{0.52}
\begin{tabular}{|l|c|c|c|c|c|l|}
\hline
Set & Dimension & \parbox{2.2cm}{RMSE [eV]} &  \parbox{2.2cm}{RMSE [eV]} &  \parbox{2.2cm}{MaxAE [eV]} &  \parbox{2.2cm}{Ratio} &  \parbox{2cm}{Descriptor} \\
 &  &  all &  CV &  CV  &  &  \\
\hline
0) all & 1 &	0.14	& 0.15	& 0.28	& 0.85 & $\mathfrak{A}$ \\
       & 2 &	0.10	& 0.11	& 0.18	& 0.99 & $\mathfrak{A,B}$ \\
       & 3 &	0.08	& 0.09	& 0.17	& 0.95 & $\mathfrak{A,B,C}$ \\
\hline
a) no CC	& 1 &	0.12	& 0.14	& 0.29	& 0.91 & $\mathfrak{D}$ \\
					& 2 &	0.10	& 0.13	& 0.26	& 0.39 & $\mathfrak{B,E}$ \\
					& 3 &	0.08	& 0.08	& 0.17	& 0.91 & $\mathfrak{D,B,C}$ \\
\hline
b) no CC, BN 	& 1 & 0.12 & 0.14	& 0.27	& 0.94 & $\mathfrak{D}$ \\
							& 2 &	0.10	& 0.12	& 0.26	& 0.36 & $\mathfrak{D,C}$ \\
							& 3 &	0.08	& 0.08	& 0.16	& 0.91 & $\mathfrak{D,C,B}$ \\
\hline
c) no BN 	& 1 &	0.14	& 0.19	& 0.39	& 0.75 & $\mathfrak{A}$ \\
					& 2 &	0.10	& 0.13	& 0.26	& 0.83 & $\mathfrak{A,B}$ \\
					& 3 &	0.08	& 0.11	& 0.24	& 0.87 & $\mathfrak{A,B,C}$ \\
\hline
d) no C & 1 &	0.12	& 0.13	& 0.27	& 0.76 & $\mathfrak{F}$ \\
				& 2 &	0.09	& 0.12	& 0.23	& 0.74 & $\mathfrak{B,E}$ \\
				& 3 &	0.07	& 0.10	& 0.20	& 0.47 & $\mathfrak{D,B,C}$ \\
\hline
e) no B & 1 &	0.13	& 0.18	& 0.37	& 0.37 & $\mathfrak{G}$ \\
				& 2 &	0.10	& 0.14	& 0.30	& 0.42 & $\mathfrak{A,B}$ \\
				& 3 &	0.08	& 0.12	& 0.24	& 0.38 & $\mathfrak{A,B,C}$ \\
\hline
f) no CdO & 1 &	0.14	& 0.18	& 0.38	& 0.77 & $\mathfrak{A}$ \\
					& 2 &	0.10	& 0.12	& 0.23	& 0.86 & $\mathfrak{A,B}$ \\
					& 3 &	0.08	& 0.10	& 0.22	& 0.90 & $\mathfrak{A,B,C}$ \\
\hline
g) no O  	& 1 &	0.13	& 0.16	& 0.33	& 0.79 & $\mathfrak{A}$ \\
					& 2 &	0.10	& 0.13	& 0.27	& 0.74 & $\mathfrak{A,B}$ \\
					& 2 &	0.08	& 0.12	& 0.25	& 0.33 & $\mathfrak{A,B,H}$ \\
\hline
h) no Cd 	& 1 &	0.14	& 0.18	& 0.38	& 0.80 & $\mathfrak{A}$ \\
					& 2 &	0.10	& 0.13	& 0.26	& 0.84 & $\mathfrak{A,B}$ \\
					& 3 &	0.08	& 0.11	& 0.24	& 0.89 & $\mathfrak{A,B,C}$ \\
\hline
\end{tabular}
\caption{Summary of the training and CV errors for 1-, 2-, 3-dimensional descriptors for different datasets. The last column reports the descriptor found using all data in the dataset for the training and the column ``Ratio'' reports the fraction of times the same descriptor was found over the L-10\%-OCV iterations. CC 
means C-diamond.\label{T:LMO}}
\end{table}
\clearpage
The code for the descriptors is:
\begin{align}
 \nonumber \mathfrak{A} &: 
\frac{\textrm{IP(B)}-\textrm{EA(B})}{r_p(\textrm{A})^2} &\, \mathfrak{B} &: 
\frac{|r_s(\textrm{A})-r_p(\textrm{B})|}{\exp(r_s(\textrm{A}))} &\, \mathfrak{C} 
&: \frac{|r_s(\textrm{B})-r_p(\textrm{B})|}{\exp(r_d(\textrm{A}))} \\
 \nonumber \mathfrak{D} &: \frac{\textrm{IP(B)}}{\exp[(r_p(\textrm{A}))^2]} &\, 
\mathfrak{E} &: \frac{|\textrm{H(B)}-\textrm{L(B)}|}{r_p(\textrm{A})^2} &\, 
\mathfrak{F} &: \frac{\textrm{H(B)}}{r_p(\textrm{A})^2}\\
\mathfrak{G} &: \frac{\textrm{IP(A)}}{[r_s(\textrm{A})+r_p(\textrm{A})]^2} &\, 
\mathfrak{H} &: \frac{\textrm{EA(A)}}{\exp[(r_d(\textrm{A}))^2]}
\end{align}
We make the following observations:
\begin{itemize}
 \item When C-diamond (and C-diamond together with BN) are excluded from the 
set, the errors are marginally smaller. However, the descriptor changes in both cases. 
In particular, the 3D descriptor is remarkably stable both for b) and a).
 \item Removal or all 4 carbon compound leads to a similar behavior as in a) and 
b). However, upon removal of BN or all boron compounds leads to a fit similar 
to 0).
 \item Carbon can be thus considered as an anomaly, but it also carries 
important information for the stability of the overall fit.
\end{itemize}

\section{\texttt{PYTHON} script for running the LASSO algorithm with 
the scikit\_learn library}
\label{A:D}

\begin{lstlisting}[language=Python]
from sklearn import linear_model
lasso = sklearn.linear_model.Lasso(alpha=lambda)
lasso.fit(D,P)
coefficients = lasso.coef_
\end{lstlisting}
Here {\tt sklearn} solves the problem in Eq. \eqref{eq:rr1}, with the 
$\ell_2$-norm scaled by a factor $\frac{1}{2N}$. The \textit{bias} $c_0$ is 
included by default.
 
\end{document}